\pdfoutput=1
\documentclass[aps,prc,reprint,showpacs,floatfix,tightenlines,superscriptaddress]{revtex4-1}

\usepackage{amsmath}
\usepackage{txfonts}
\usepackage{graphicx}
\usepackage{color}

\usepackage[colorlinks=true,linkcolor=blue,citecolor=blue,urlcolor=blue]{hyperref}

\hypersetup{
  pdftitle={Short-range correlations in nuclei with similarity renormalization group transformations},%
  pdfauthor={Thomas Neff <t.neff@gsi.de>}
}

\renewcommand{\vec}[1]{\boldsymbol{\mathbf{#1}}}
\newcommand{\op}[1]{\widehat{#1}}
\newcommand{\conj}[1]{{#1}^*}
\newcommand{\adj}[1]{{#1}^\dagger}

\newcommand{\comm}[2]{\bigl[ {#1}, {#2} \bigr]}

\newcommand{\bra}[1]{\bigl\langle {#1} \bigr\rvert}
\newcommand{\ket}[1]{\bigl\vert {#1} \bigr\rangle }
 
\newcommand{\matrixe}[3]{\bigl\langle {#1} \big\vert {#2} \big\vert {#3} \bigr\rangle}

\newcommand{\cg}[6]{\left\langle \begin{matrix} {#1} & {#3} \\ {#2} & {#4} \end{matrix}\, 
\right\rvert\left. \begin{matrix} {#5} \\ {#6} \end{matrix} \right\rangle}

\newcommand{\talmi}[6]{\left\langle {#1},{#2} : {#5} \right\vert \left. {#3},{#4} : {#5} \right\rangle_{#6}}

\newcommand{\fm}{\ensuremath{\textrm{fm}}}
\newcommand{\MeV}{\ensuremath{\textrm{MeV}}}
\newcommand{\nuc}[2]{$^{#1}${#2}}

\DeclareMathOperator{\Tr}{Tr}

\newcommand{\scalefig}{0.90}

\begin{document}

\title{Short-range correlations in nuclei with similarity renormalization group transformations} 

\author{T. Neff}
\email{email: t.neff@gsi.de}
\affiliation{GSI Helmholtzzentrum f\"ur Schwerionenforschung GmbH,
Planckstra{\ss}e~1, 64291~Darmstadt, Germany}

\author{H. Feldmeier}
\affiliation{GSI Helmholtzzentrum f\"ur Schwerionenforschung GmbH,
Planckstra{\ss}e~1, 64291~Darmstadt, Germany}
\affiliation{Frankfurt Institute for Advanced Studies, Max-von-Laue
Stra{\ss}e~1, 60438~Frankfurt, Germany}

\author{W. Horiuchi}
\affiliation{Department of Physics, Hokkaido University, Sapporo 060-0810, Japan}


\begin{abstract} 
  \begin{description}
   \item[Background]
   	Realistic nucleon-nucleon interactions induce short-range correlations
		in nuclei. To solve the many-body problem unitary transformations like
		the similarity renormalization group (SRG) are often used to soften the interactions.
   \item[Purpose] 
   	Two-body densities can be used to illustrate how the SRG eliminates
		short-range correlations in the wave function. The short-range information 
		can however be recovered by transforming the density operators.
   \item[Method] 
    The many-body problem is solved for \nuc{4}{He} in the no
    core shell model (NCSM) with SRG transformed AV8' and chiral N3LO interactions. 
   	The NCSM wave functions are used to calculate two-body densities with
		bare and SRG transformed density operators in two-body approximation.
   \item[Results]
   	The two-body momentum distributions for AV8' and N3LO have
		similar high-momentum components up to relative momenta of about $2.5\,\fm^{-1}$,		
		dominated by tensor correlations, but differ in their behavior at higher relative
		momenta. The contributions of many-body correlations are small for pairs
		with vanishing pair momentum but not negligible for the momentum distributions
		integrated over all pair momenta. Many-body correlations are induced by
		the strong tensor force and lead to a reshuffling of pairs between
		different spin-isospin channels.		   
   \item[Conclusions] 
  	When using the SRG it is essential to use transformed operators for 
		observables sensitive to short-range physics. Back-to-back pairs with
		vanishing pair momentum are the best tool to study short-range
		correlations.    
  \end{description} 
\end{abstract}


\maketitle

\section{Introduction}
\label{sec:intro}

Realistic nucleon-nucleon ($NN$) interactions are fitted to $NN$ scattering data up
to the pion production threshold. Therefore their properties at short distances $r$ 
and also their off-shell behavior is not completely constrained. All realistic
interactions include pions to describe the long- and medium-range parts of the
potential. The short-range part is parameterized phenomenologically \cite{wiringa95},
by the exchange of heavy mesons \cite{machleidt01}, or in chiral effective field theory by
(regularized) contact-terms \cite{entem03,epelbaum05}. At short distances $r \lesssim 1.5\,\fm$ the
different interactions induce typical short-range correlations due to short-range repulsion
and the tensor force which is reflected in a depopulation of one-body momentum distributions
below the Fermi momentum and an enhancement at high momenta when compared to single-particle
mean-field occupation probabilities \cite{benhar94,pandharipande97,muether00,ucom03,dickhoff04,rios09}.

However short-range correlations are two-body correlations and therefore two-body densities
in coordinate and momentum space provide the best tool to study these
correlations. Experimentally short-range correlations have been studied in inclusive, semi-inclusive
and triple-coincidence reactions, see the reviews \cite{frankfurt08,arrington12} and references therein. 
The most direct information about two-body correlations can be obtained in 
triple-coincidence experiments where one knocks out a nucleon pair with protons \cite{tang03} or electrons
\cite{shneor07,subedi08,badhdasaryan10,korover14} at high-momentum transfer. In these experiments
one found a dominance of $pn$ over $pp$ pairs at high relative momenta, that indicated the
importance of short-range tensor correlations. Tensor correlations also appear to play a
major role in $(p,d)$ reactions at high proton energies \cite{miki13,ong13}.

Recent theoretical studies of two-body momentum distributions can explain these
observations. In a simple picture only close nucleons found in relative $S$-wave pairs are affected
by short-range correlations and in a first approximation high-momentum components
are generated by pairs in deuteron-like configurations \cite{vanhalst12}. The two-body momentum
distributions at high momenta can also be connected to the nuclear contacts \cite{weiss15}.
Detailed few- and many-body calculations almost exclusively use the 
Argonne~$v_{18}$ (AV18) or Argonne~$v_8'$ (AV8') interactions \cite{schiavilla07,wiringa08,wiringa14,alvioli12,alvioli13a}. 
Compared to the Argonne interactions, interactions derived in chiral effective field theory \cite{entem03} are
regularized with a relatively low momentum cut-off and one might expect noticeable differences for the short-range correlations.

From the perspective of nuclear many-body calculations short-range correlations
are not a desired feature but pose a severe problem. One way to address this problem is to include
the short-range correlations explicitly, as in the correlated basis function 
theory \cite{fabrocini00} or in variational Monte Carlo (VMC) calculations \cite{schiavilla07,wiringa08,wiringa14}.
Another approach is to use soft phase-shift equivalent effective interactions that are obtained from the
bare interactions by means of unitary transformations like $V_{\mathit{low-k}}$ \cite{bogner03},
the unitary correlation operator method (UCOM) \cite{ucom98,ucom03,ucom10} or
the similarity renormalization group (SRG) \cite{bogner07,bogner10,ucom10}. 
In real life the unitary transformations are performed in an $n$-body
approximation. This usually means in two-body, or for state of the art SRG calculations,
in three-body approximation \cite{jurgenson09,hebeler12,roth14}. 
Even if we only have a two-body interaction at the beginning the unitary transformation 
will induce three- and higher-body terms.  
If the unitary transformation acts mainly at short distances and the density of the
nuclear system is low enough, so that the probability to find three nucleons
simultaneously close together is small, transformations on the two-body level
will be a good approximation. The remaining dependence of observables on the
unitary transformation can be used to analyze the nature of missing
many-body terms.

In a consistent calculation it is not enough to transform the Hamiltonian, all operators
have to be transformed. Within the SRG approach the transformation of long-range operators like radius
or electromagnetic transition operators has been studied in \cite{schuster14} in two- and
three-body approximation. As one might expect the effect on these long-range operators
is not very large. On the other hand we expect large effects for short-range or
high-momentum observables. Within the UCOM approach we investigated
these effects in two-body approximation and with simple trial wave functions
for the one-body momentum distribution \cite{ucom03} and for two-body coordinate and
momentum space distributions \cite{src11}. The SRG operator evolution for the deuteron was studied 
extensively in \cite{anderson10} and analyzed in terms of a factorization for high-momentum
observables. These ideas were extended for Fermi gases in \cite{bogner12}.

In this paper we investigate two-body densities for the ground state of
\nuc{4}{He} in coordinate and momentum space for SRG transformed AV8'
\cite{wiringa95} and chiral N3LO \cite{entem03} interactions using the no core shell model (NCSM).
For the bare AV8' interaction the NCSM results are not converged and we use in this case the method of
correlated Gaussians \cite{src11,suzuki09,mitroy13}. We will discuss the two-body densities 
obtained with both, bare and transformed density operators. The SRG transformation 
depends on the flow parameter that controls the softness of the transformed 
interaction. We perform the SRG transformation in two-body approximation and use the
flow-dependence of the results to test the quality of the two-body approximation.
We will show that for pairs with vanishing pair momentum the contributions of
three-body correlations are negligible. 

After describing in Sec.~\ref{sec:method} the used methods, two-body
densities in coordinate and momentum space are presented 
in Sec.~\ref{sec:results} where we also discuss the role of many-body correlations
for different observables. Summary and conclusions follow in Sec.~\ref{sec:summary}.
Technical details about the calculation of translational invariant two-body densities
in the NCSM framework are presented in the Appendix.

\section{Method}
\label{sec:method}

The SRG flow equation for the Hamiltonian
$\op{H}_\alpha = \op{T}_{\mathrm{int}}+\op{V}_\alpha$ and the corresponding
transformation matrix $\op{U}_\alpha$ are given by 
\begin{equation} \label{eq:evol}
  \frac{d\op{H}_\alpha}{d\alpha} = \comm{\op{\eta}_\alpha}{\op{H}_\alpha}, \quad
  \frac{d\op{U}_\alpha}{d\alpha} = -\op{U}_\alpha  \op{\eta}_\alpha  \: ,
  \end{equation}
with the generator $\op{\eta}_\alpha$ taken to be the commutator of the
intrinsic kinetic energy and the evolved Hamiltonian:
\begin{equation}
  \op{\eta}_\alpha = (2\mu)^2\; \comm{\op{T}_{\text{int}}}{\op{H}_\alpha}, \quad
  \op{T}_{\text{int}}=\op{T}-\op{T}_{\text{cm}} \: .
\end{equation}
Please note that $\op{U}_\alpha$ used here and in Ref.~\cite{ucom10} corresponds
to $\adj{\op{U}_s}$ in Refs.~\cite{bogner07,bogner10}.

In this paper the Hamiltonian is transformed in two-body approximation:
\begin{equation} \label{eq:2-bodyapp}
  \op{H}_\alpha = \adj{\op{U}_\alpha} \op{H} \op{U}_\alpha = 
  \op{T}_\mathrm{int} + \op{V}_\alpha^{[2]} + \ldots + \op{V}_\alpha^{[N]} 
  \approx \op{T}_\mathrm{int} + \op{V}_\alpha^{[2]} \: .
\end{equation}
The evolution \eqref{eq:evol} can therefore be performed for the relative motion
in two-body space. The flow equations are solved on a momentum space grid with relative momenta
going up to $k_\mathrm{max} = 15\,\fm^{-1}$ for the AV8' interaction.
In the end the momentum space matrix elements are evaluated in the harmonic oscillator
basis to be used in the NCSM. We will present results for flow parameters of $\alpha = 0.01\,\fm^4$, 
$\alpha = 0.04\,\fm^4$ (a typical value used in many-body calculations), and $\alpha = 0.20\,\fm^4$
corresponding to a very soft effective Hamiltonian $\op{H}_\alpha$.

The many-body problem for \nuc{4}{He} is then solved with the SRG transformed two-body
Hamiltonian $\op{H}_\alpha$ using the shell model code \textsc{Antoine}
\cite{caurier99}:
\begin{equation}
 \op{H}_\alpha \ket{\Psi_\alpha} = E_\alpha \ket{\Psi_\alpha} \: .
\end{equation}
The results for the ground state energy of \nuc{4}{He} and the two-body distributions
in coordinate and momentum space discussed in this paper are well converged within the
model space ($N_\mathrm{max} = 16$, oscillator parameter $\hbar\Omega = 36
\MeV$). The bare AV8' interaction ($\alpha=0$) can however not be converged in
this space and we use here the results obtained with the correlated Gaussian
method \cite{src11,suzuki09,mitroy13}.

All two-body information contained in the many-body states $\ket{\Psi_\alpha}$
can be expressed in terms of the two-body density matrix
that is obtained by integrating over all coordinates besides the pair position
and the relative coordinates of the pair. The two-body density as calculated in
the NCSM is not translationally invariant due to the localization of the NCSM
wave function in the origin of the coordinate system. However the translational
invariant two-body density can be obtained from the two-body density in the
laboratory system by a linear transformation as discussed in 
Appendix~\ref{app:two-body-density-intrinsic}. In the harmonic oscillator basis
we express the two-body density matrix $\rho^{\alpha}_{qQ;q'Q'}$ with two-body basis states
\begin{equation}
	\begin{split}
	  \ket{qQ} & = \ket{q} \otimes \ket{Q} \\
  	&= \ket{b_\mathrm{rel}; nlm, S M_S, T M_T} \otimes \ket{b_\mathrm{pair}; NLM} \: .
	\end{split}
\end{equation}
Here $q$ summarizes the harmonic oscillator quantum numbers. $n l m$ stand for the
relative coordinates, $SM_S$ and $TM_T$ give the total spin and isospin of the pair,
respectively, and $Q$ summarizes the harmonic oscillator quantum numbers for
the pair coordinate. The oscillator parameters for the relative and the pair motion
are given by
\begin{equation}
	b_\mathrm{rel} = \sqrt{2}\,b, \quad 
	b_\mathrm{pair} = \sqrt{\frac{A}{2(A-2)}}\, b \: ,
\end{equation}
respectively.

A compact notation can be obtained by defining the two-body 
density operator, which acts in two-body space, as
\begin{equation}
  \op{R}_{\alpha} = \sum_{qQ,q'Q'} \ket{qQ}\,\rho^{\alpha}_{qQ;q'Q'}\,\bra{q'Q'} \: .
\end{equation}

The expectation value of any `bare' two-body operator $\op{B}$ can then
simply evaluated as
\begin{equation} \label{eq:2-body-bare} 
  B_\alpha := \matrixe{\Psi_\alpha}{\op{B}}{\Psi_\alpha} 
  = \Tr_2 \left( \op{R}_{\alpha}\op{B} \right) \: ,
\end{equation}
where $\Tr_2$ denotes the trace in two-body space. If one evolves the observable
$\op{B}$ in the same way as the Hamiltonian, namely
$\op{B}_\alpha = \adj{\op{U}_\alpha} \op{B} \op{U}_\alpha$, the expectation
value
\begin{equation}
	\begin{split}  
    \matrixe{\Psi_\alpha}{\op{B}_\alpha}{\Psi_\alpha} &= 
    \matrixe{\Psi_\alpha}{\adj{\op{U}_\alpha} \op{B} \op{U}_\alpha}{\Psi_\alpha} \\
  	&= \matrixe{\Psi_{\alpha=0}}{ \op{B} }{\Psi_{\alpha=0}}
	\end{split}
\end{equation}
does not depend on $\alpha$ because $U_\alpha \ket{\Psi_\alpha} = \ket{\Psi_{\alpha=0}}$. 
However, as for the Hamiltonian, the evolved observable $\op{B}_\alpha$ is in general 
no longer a two-body operator and contains induced higher-body operators.

The two-body approximation consists in calculating
\begin{equation} \label{eq:2-body-approx} 
  \widetilde{B}_\alpha := \Tr_2 \left( \op{R}_{\alpha}\,\op{B}_\alpha \right)
  = \Tr_2 \left( \op{U}_\alpha \op{R}_{\alpha} \adj{\op{U}_\alpha}\,\op{B} \right) \: ,
\end{equation}
where the two-body operator $\op{B}_\alpha$ is SRG transformed in two-body
space. It should be noted that $\op{U}_\alpha$ acts only on the
relative coordinate part of $\op{R}_{\alpha}$ or $\op{B}$ and leaves the
center of mass motion of the pair unchanged.

If the two-body approximation was exact (for both the Hamiltonian $\op{H}$ and the observables $\op{B}$) 
$\widetilde{B}_\alpha$ would be $\alpha$-independent. The remaining $\alpha$ dependence 
therefore indicates the size of the neglected contributions from the induced three- and higher-body terms. 

With the many-body states $\ket{\Psi_\alpha}$ obtained in the NCSM
various bare and SRG transformed two-body quantities will be studied
in the following. With bare operators we can investigate how the properties of
the eigenstates $\ket{\Psi_\alpha}$ change with increasing flow parameter. This
will reflect the increasing `softness' of the transformed Hamiltonian $\op{H}_\alpha$.
On the other hand properties calculated with transformed operators should be independent
from the flow parameter if the two-body approximation is justified. However one has to be
careful when drawing conclusions about omitted higher-order terms in the transformed
operators. In this paper the two-body approximation is employed both for the transformed
Hamiltonian and for the transformed operators. Even if the two-body approximation is perfect
for the transformed operators we would still expect a dependence on the flow parameter due
to the two-body approximation for the transformation of the Hamiltonian. 

It might be possible however that particular components of the wave function are less sensitive
to higher-order terms in the transformed Hamiltonian and can be obtained reliably within the 
two-body approximation. This appears to be the case for the pairs with small pair momentum as
will be discussed in Sec.~\ref{sec:vanishing-pair-momentum}.

\section{Results}
\label{sec:results}

\subsection{Energies}
\label{sec:energies}

\begin{figure}
  \centering
  \includegraphics[width=0.65\columnwidth]{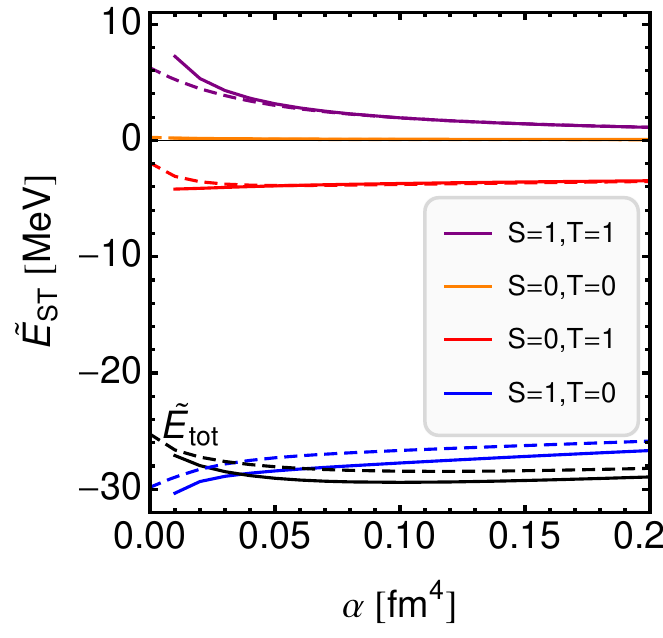}
  \caption{(Color online) Contributions from the different $S,T$ channels to the
  total energy of \nuc{4}{He} as a function of the flow parameter $\alpha$ for
  the AV8' interaction (solid) and the N3LO interaction (dashed). The total energy $\tilde{E}_\mathrm{tot}$
  (black lines) includes the Coulomb energy.}
  \label{fig:energy-contributions}
\end{figure}

In two-body approximation $\op{U}_\alpha$ is exactly unitary in two-body
space, but in many-body space unitarity is only approximate. Therefore the energy
eigenvalue $E_\alpha$ depends on $\alpha$ and this dependence can be taken as
a measure for the induced three- and four-body interactions (see
Eq.~\eqref{eq:2-bodyapp}) that are neglected when calculating the \nuc{4}{He}
ground state energy in two-body approximation. As seen in
Fig.~\ref{fig:energy-contributions} the total energy varies by about 10\%
indicating that the contribution of three- and four-body terms are small.
In Fig. \ref{fig:energy-contributions} we also show the individual contributions from the four $S,T$
channels to the total ground state energy of \nuc{4}{He}. These can be calculated by 
using Eq.~\eqref{eq:2-body-approx}
\begin{equation}
	\tilde{E}_{\alpha,ST} = \Tr_2 \left( \op{R}_{\alpha} 
	\bigl[ (\op{T}_\mathrm{int} + \op{V}_{\alpha}^{[2]}) \op{\Pi}_{ST} \otimes \op{1} \bigr] \right) \: ,
\end{equation}
where the operator $\op{\Pi}_{ST}$ projects on the spin-isospin channel $S,T$.
The total energies for the AV8' and N3LO interactions are very similar (although the individual
kinetic and potential contributions are quite different) and in both cases
the dominant contribution to the binding energy is coming from $S,T = 1,0$ pairs with
a large contribution from the tensor force. The $S,T=0,1$ channel gives a much smaller attractive
contribution whereas the $S,T=1,1$ channel provides repulsion and the contribution from the $S,T=0,0$
channel is negligible. With increasing flow parameter we observe a reduction of the attractive
contribution of the $S,T=1,0$ channel and the $S,T=1,1$ contribution becomes less repulsive. These 
changes are related to changes in the occupation of the different $S,T$ channels that will be
discussed in Sec.~\ref{sec:st-rel-mom}.

\subsection{Relative density distribution in coordinate space}
\label{sec:rel-dens}

The repulsive nature of the nuclear interaction is easily seen in the 
the relative density distribution $\rho^\mathrm{rel}_\alpha(r)$ that gives
the probability to find a pair of nucleons
at a distance $r$. It is calculated by replacing $\op{B}$ in
Eq.~\eqref{eq:2-body-bare} with the two-body operator 
\begin{multline}
 \op{\rho}^\mathrm{rel}(r) = \\
 \sum_{l m S M_S T M_T} \ket{r l m, S M_S,T M_T} \bra{r l m, S M_S,T M_T} \otimes \op{1} \: ,
\end{multline}
where we sum over all quantum numbers except the radial distance $r$. 

\begin{figure}
  \centering
  \includegraphics[width=\columnwidth]{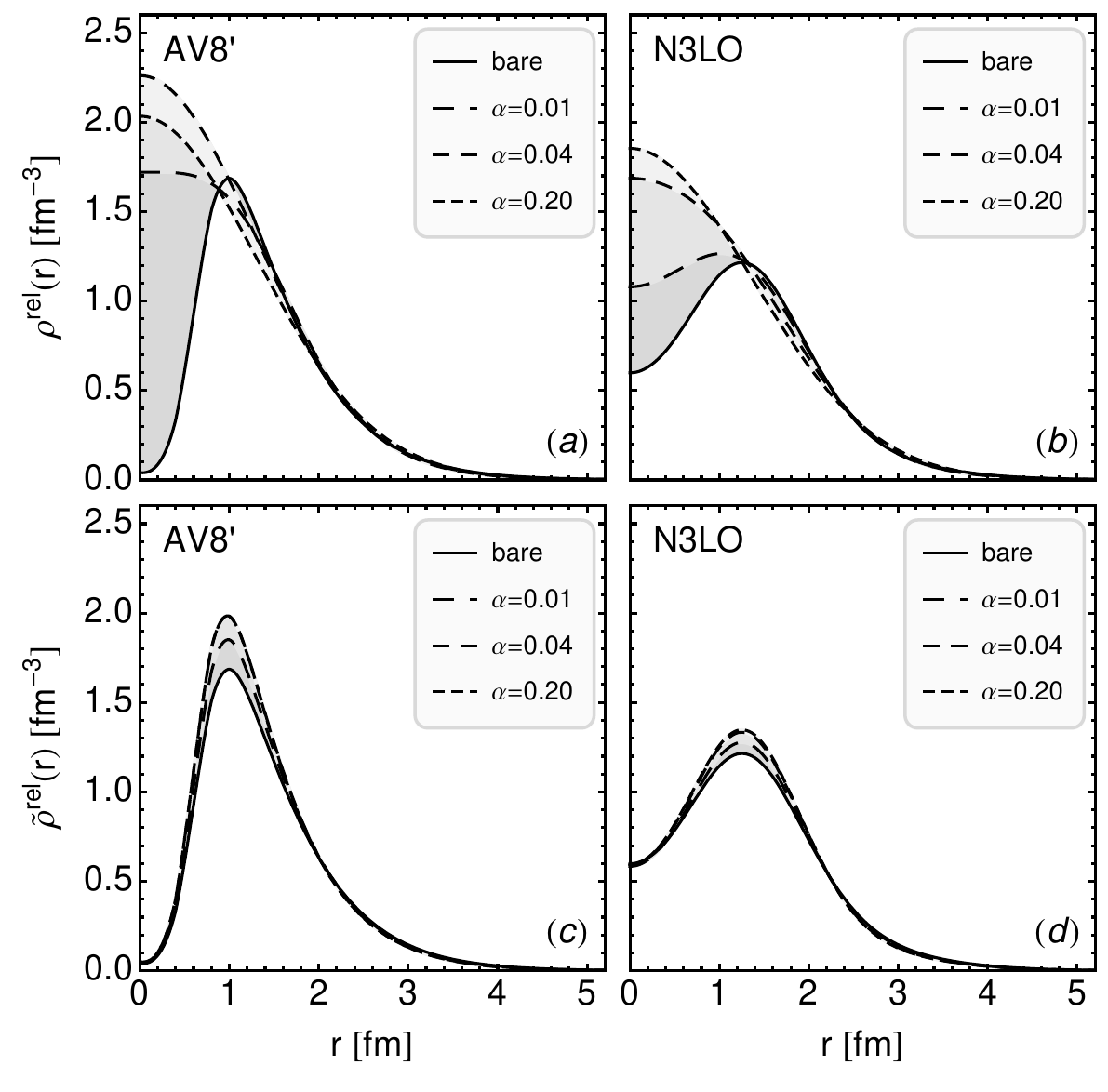}
  \caption{AV8' (left) and N3LO (right) two-body densities in coordinate space calculated 
  with bare (top) and SRG transformed density operators (bottom). $\alpha$ in units of $\fm^4$.
  See also Fig.~1 in Ref.~\cite{src15} for the AV8' two-body densities in the different $S,T$
  channels.}
  \label{fig:tbdens-coordinate}
\end{figure}

The short-range correlations induced by the nuclear interaction are reflected in the
relative density distributions $\rho^\mathrm{rel}_\alpha(r)$ shown in the upper part of 
Fig.~\ref{fig:tbdens-coordinate}. They are calculated with the eigenstates of the bare
and SRG evolved Hamiltonians $\op{H}_\alpha$ according to Eq.~\eqref{eq:2-body-bare}. 
For the bare interactions the relative density distributions
$\rho^\mathrm{rel}_{\alpha=0}(r)$ show the typical correlation hole at short
distances. Due to the repulsive core of the interaction the probability to find a pair of nucleons at a
distance $r$ between their centers is depleted at distances below $r \approx 1\,\fm$. 
The chiral N3LO interaction is obviously not as repulsive as the AV8' interaction at short
distances, as the correlation hole is less pronounced for the N3LO interaction. 

One should keep in mind that the term `correlation hole' is misleading
as for $r \approx 0.5\,\fm$ the nucleon densities are already strongly
overlapping so that there is no hole in the baryonic matter density. On the
contrary at short distances one encounters locally large baryonic matter
densities and expects strongly polarized nucleons. 

With increasing flow parameter $\alpha$ the correlation hole disappears more and
more. Furthermore the density distributions calculated for the evolved AV8' and
N3LO interactions become increasingly similar, cf. Fig.~\ref{fig:tbdens-coordinate}
(a), (b). For the largest flow parameter the relative density distribution is
essentially of Gaussian shape --- as would be expected for an uncorrelated
mean-field wave function. Thus, the SRG transformation brings us from a highly correlated
system to a simple shell model or mean-field like situation.

On the other hand the density distributions
$\tilde{\rho}^\mathrm{rel}_\alpha(r)$ shown in the lower part of
Fig.~\ref{fig:tbdens-coordinate}, which are obtained with the SRG transformed
density operators according to Eq.~\eqref{eq:2-body-approx}, are all very
similar to those obtained for the bare Hamiltonians
$\op{H} \equiv \op{H}_{\alpha=0}$. Without employing the two-body approximation $\op{U}_\alpha$
would be unitary in four-body space and $\tilde{\rho}_\alpha$ would not depend
on $\alpha$. The remaining $\alpha$ dependence indicates that the induced three-
and four-body terms are small but not completely negligible.

The probability densities obtained with both bare and transformed density operators 
are normalized to the number of pairs:
\begin{equation}
 \label{eq:coord-normalization}
 \int_0^\infty dr \: r^2 \: \rho^\mathrm{rel}_\alpha(r) = 
 \int_0^\infty dr \: r^2 \: \tilde{\rho}^\mathrm{rel}_\alpha(r) = \frac{A(A-1)}{2}
\end{equation}

The spatial distributions shown in Fig.~\ref{fig:tbdens-coordinate} may help our
intuition but can not easily be related to experiment. Therefore momentum
distributions and momentum correlations will be addressed in the following.

\subsection{Relative momentum distributions}
\label{sec:rel-mom}

The relative momentum distribution, i.e., the probability $n^\mathrm{rel}_{\alpha,lST}(k)$ to find a
nucleon pair with relative momentum $k$, relative orbital angular momentum $l$, spin $S$, and 
isospin $T$ is obtained by replacing $\op{B}$ in Eq.~\eqref{eq:2-body-bare} with the two-body operator 
\begin{multline}\label{eq:tbdens-momentum-st}
 \op{n}^\mathrm{rel}_{lST}(k) = \\
 \sum_{m M_S M_T} \ket{k l m, SM_S, TM_T} \bra{k l m, SM_S, TM_T} \otimes \op{1} \: ,
\end{multline}
where $\ket{k l m, SM_S, TM_T}$ denotes the spherical momentum space
representation of the relative motion. 

\begin{figure}
  \centering
  \includegraphics[width=\columnwidth]{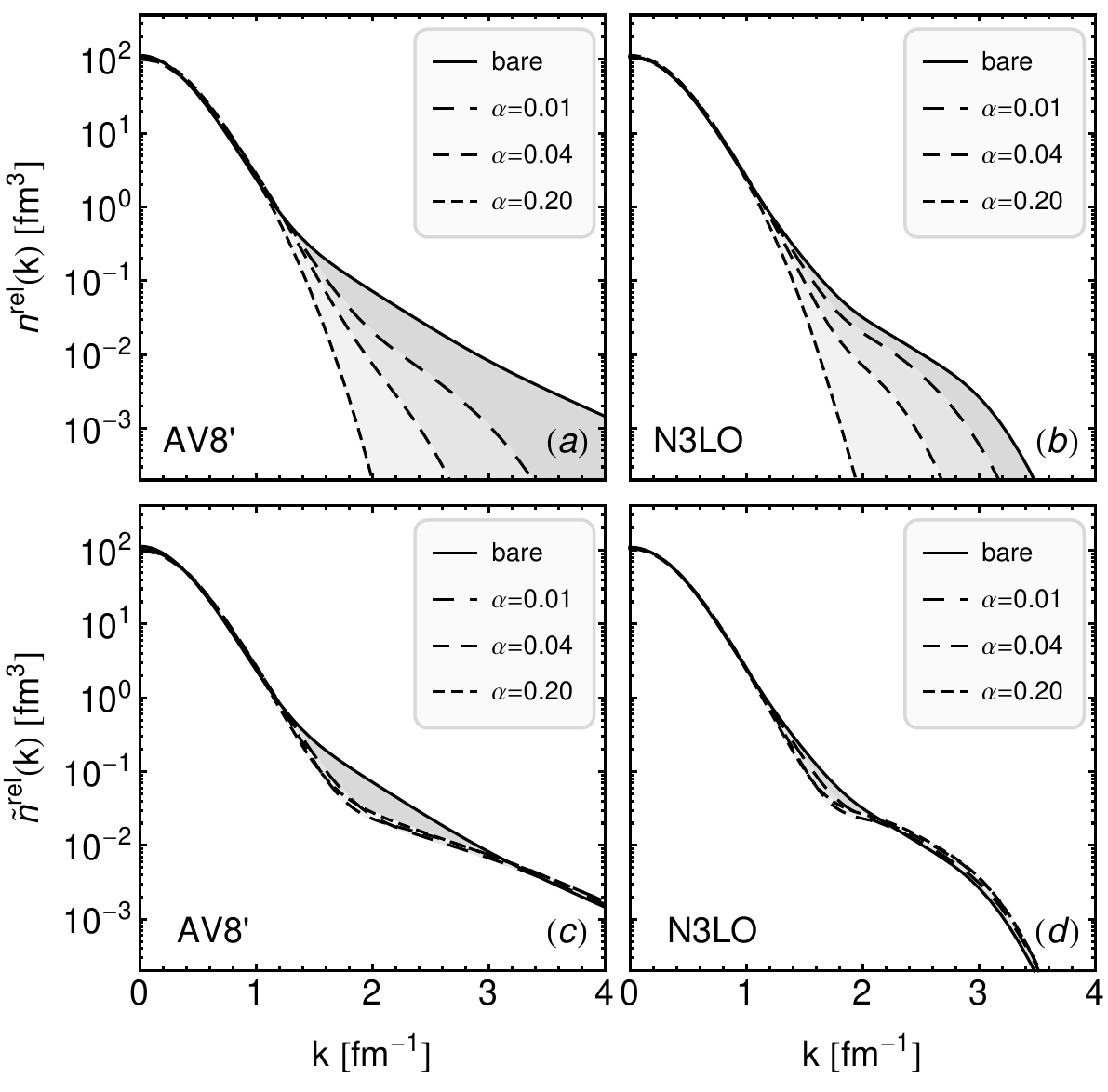}
  \caption{AV8' (left) and N3LO (right) two-body densities in momentum-space calculated with bare (top) 
  and SRG transformed density operators (bottom). $\alpha$ in units of $\fm^4$. See also Fig.~2 in
  Ref.~\cite{src15} for the AV8' two-body densities in the different $S,T$ channels.} 
  \label{fig:tbdens-momentum}
\end{figure}

The relative momentum distributions for pairs with relative momentum $k$, irrespective of
their orbital angular momentum $l$, spin $S$ and isospin $T$,
$n^\mathrm{rel}_{\alpha}(k) = \sum_{lST} n^\mathrm{rel}_{\alpha,lST}(k)$, and its SRG
transformed partner $\tilde{n}^\mathrm{rel}_{\alpha}(k)$ are shown in
Fig.~\ref{fig:tbdens-momentum} for the two interactions and different flow
parameters. 
The short-range repulsive correlations, which manifest themselves in coordinate
space as the correlation holes, show up as tails in the momentum distributions 
$n^\mathrm{rel}_\alpha(k)$ that reach out to large relative momenta $k$. Note however
that the momentum distribution is not the Fourier transform of the diagonal two-body density in
coordinate space.

Whereas the tail
of the momentum distribution shows an exponential behavior at large relative momenta for
the AV8' interaction, the N3LO relative momentum distribution reflects the
momentum space regulator that cuts off high momenta beyond about $3.5\,\fm^{-1}$.
For both interactions short-range tensor correlations play an important role as
will be discussed in more detail later. 
With increasing flow parameter $\alpha$ the high-momentum components are more and more 
reduced until the probability distribution of the relative momentum assumes a Gaussian 
shape for both evolved interactions, cf. Fig.~\ref{fig:tbdens-momentum} (a) and (b), 
corresponding to an uncorrelated wave function. 

Fig.~\ref{fig:tbdens-momentum} (c) and (d) show that the density distributions
$\tilde{n}^\mathrm{rel}_\alpha(k)$ obtained with the SRG transformed density
operators are again all very similar to those obtained for the bare Hamiltonian,
indicating that the induced three- and four-body terms are also small in
momentum space. The most visible dependence on $\alpha$ occurs for the AV8'
interaction around $k\approx 2\,\fm^{-1}$. This dependence is related to three-body correlations
induced by the two-body tensor force to be discussed in the following
subsection.

\subsection{$S,T$ relative momentum distributions}
\label{sec:st-rel-mom}

\begin{figure*}
  \centering
  \includegraphics[width=\scalefig\textwidth]{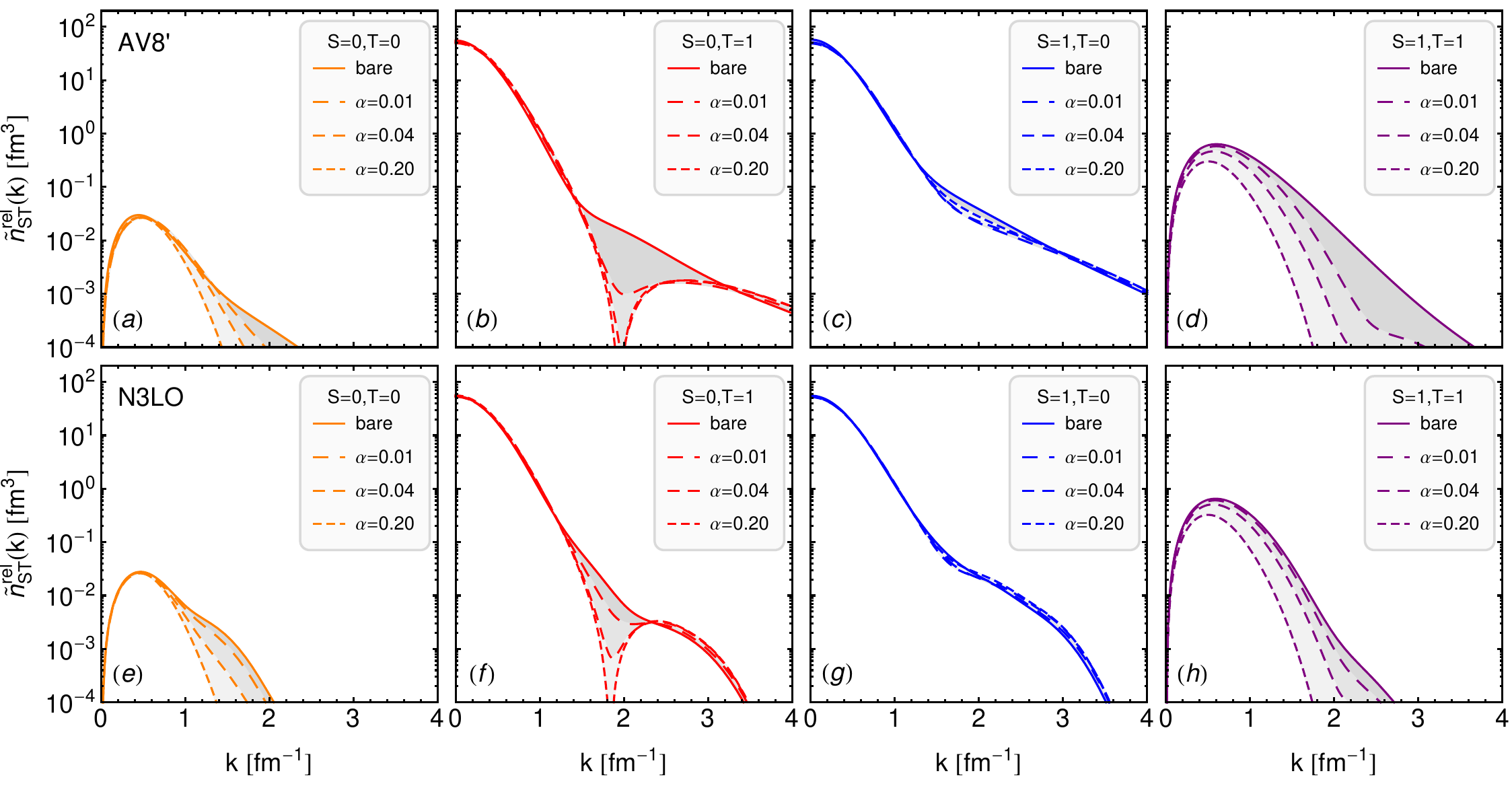}
  \caption{(Color online) AV8' (top) and N3LO (bottom) relative momentum distributions in the different 
  spin-isospin channels obtained with SRG transformed density operators. $\alpha$ in units of $\fm^4$.} 
  \label{fig:tbdens-momentum-st}
\end{figure*}

In order to get a deeper understanding of the nature of the correlations we
separate the momentum distribution according to Eq.~\eqref{eq:tbdens-momentum-st} 
into its parts $\tilde{n}^\mathrm{rel}_{\alpha,ST}(k) = \sum_l \tilde{n}^\mathrm{rel}_{\alpha,lST}(k)$ 
coming from the different $S,T$ channels
and display the results in Fig.~\ref{fig:tbdens-momentum-st}. The first observation
is that the $S,T = 1,0$ channel, in which the tensor force is strongest, shows
only a weak $\alpha$ dependence. In contrast to that there is strong $\alpha$ dependence
in the $S,T=0,1$ channel for momenta around $2\,\fm^{-1}$ and also in the $S,T=1,1$ channel.
With increasing $\alpha$ strength moves from the odd $S,T = 1,1$ to the even $S,T = 0,1$
channel. This effect has been discussed in detail in Ref.~\cite{src11}. In a simplified picture
one may consider the situation of a localized $S,T = 0,1$ pair with a third nucleon of different flavor not
too far away. Additional binding from the strong tensor force acting between the third nucleon
and a nucleon of the pair can be obtained by flipping the spin of one nucleon in the $S,T = 0,1$
pair and thus converting with some probability the $S,T=0,1$ pair into an  $S,T = 1,1$ pair. 
It is energetically more favorable to break some $S,T = 0,1$ pairs and loose their binding 
if more binding from the tensor interaction with the third particle can be gained. These genuine three-body
correlations are lost in two-body approximation and introduce an $\alpha$-dependence.
With increasing $\alpha$ the tensor part of the SRG transformed Hamiltonian $\op{H}_\alpha$
is weakened while the central part is strengthened. With the weakening of the tensor force the effect
of the three-body correlations will be reduced. This explains the reduction of the number of $S,T = 1,1$ pairs with increasing
$\alpha$ and the $\alpha$-dependence of the $S,T = 0,1$ pairs at momenta around $2\,\fm^{-1}$
where the relative importance of the tensor force is most pronounced.

\begin{table}
  \caption{Number of pairs in the different $(S,T)$ channels for the bare and SRG evolved AV8' and N3LO 
interactions.}
  \label{tab:pairs}
\begin{ruledtabular}
\begin{tabular}{l|cccc}
$n^\mathrm{pair}_{\alpha,ST}=\tilde{n}^\mathrm{pair}_{\alpha,ST}$ & (0,0) & (0,1) & (1,0) & (1,1) \\[1mm] 
\hline
AV8', bare & 0.008 & 2.572 & 2.992 & 0.428\\
AV8', $\alpha = 0.01\,\fm^4$ & 0.008 & 2.708 & 2.992 & 0.292\\
AV8', $\alpha = 0.04\,\fm^4$ & 0.007 & 2.821 & 2.993 & 0.179\\
AV8', $\alpha = 0.20\,\fm^4$ & 0.005 & 2.925 & 2.995 & 0.075\\\hline
N3LO, bare & 0.009 & 2.710 & 2.991 & 0.290\\
N3LO, $\alpha = 0.01\,\fm^4$ & 0.007 & 2.745 & 2.992 & 0.255\\
N3LO, $\alpha = 0.04\,\fm^4$ & 0.006 & 2.817 & 2.994 & 0.183\\
N3LO, $\alpha = 0.20\,\fm^4$ & 0.004 & 2.921 & 2.995 & 0.079
\end{tabular}
\end{ruledtabular}
\end{table}

This transfer of probability between the even $S,T = 0,1$ and the odd $S,T = 1,1$ 
channel can be seen in Table~\ref{tab:pairs} where the number of pairs,
$\tilde{n}^\mathrm{pair}_{\alpha,ST}$ given by the integrals over the momentum
distributions, are listed as a function of $\alpha$. With increasing flow parameter
the occupation numbers approach the limit of the mean-field or independent particle model
with 3 pairs in both even channels and 0 pairs in the odd channels. The number of pairs
$n^\mathrm{pair}_{\alpha,ST}$ and $\tilde{n}^\mathrm{pair}_{\alpha,ST}$ obtained with bare
and SRG transformed density operators are identical as
$\op{U}_\alpha$ in two-body approximation does not connect different $S,T$ channels.

The reshuffling of probability between the different spin-isospin channels also
tells us that the omitted many-body terms in $\op{U}_\alpha$ and the transformed Hamiltonian 
$\op{H}_\alpha$ will have a non-trivial spin- and isospin-dependence.

\subsection{Relative momentum distributions for $K=0$ pairs}
\label{sec:vanishing-pair-momentum}
 
Up to now we have investigated relative momentum distributions for all pairs, indiscriminate of
the pair momentum $K$. It has been found that the relative momentum distributions depend
quite significantly on the pair momentum \cite{wiringa08,alvioli13}. In the context of this
paper it is interesting to see how this is related to many-body correlations. 

We might expect that back-to-back pairs with $\vec{K} \approx 0$ are less affected by many-body
correlations than pairs with a large pair momentum $\vec{K}$. In a $K=0$ pair with large
relative momentum $\vec{k}$ both nucleons have large individual momenta. For pairs with large
pair momentum $\vec{K}$ however there is a high probability that one of the
nucleons has a momentum less than or close to Fermi momentum. We would therefore expect that these
nucleons are interacting more strongly with other nucleons and therefore are susceptible to
many-body correlations.

\begin{figure*}
  \centering
  \includegraphics[width=\scalefig\textwidth]{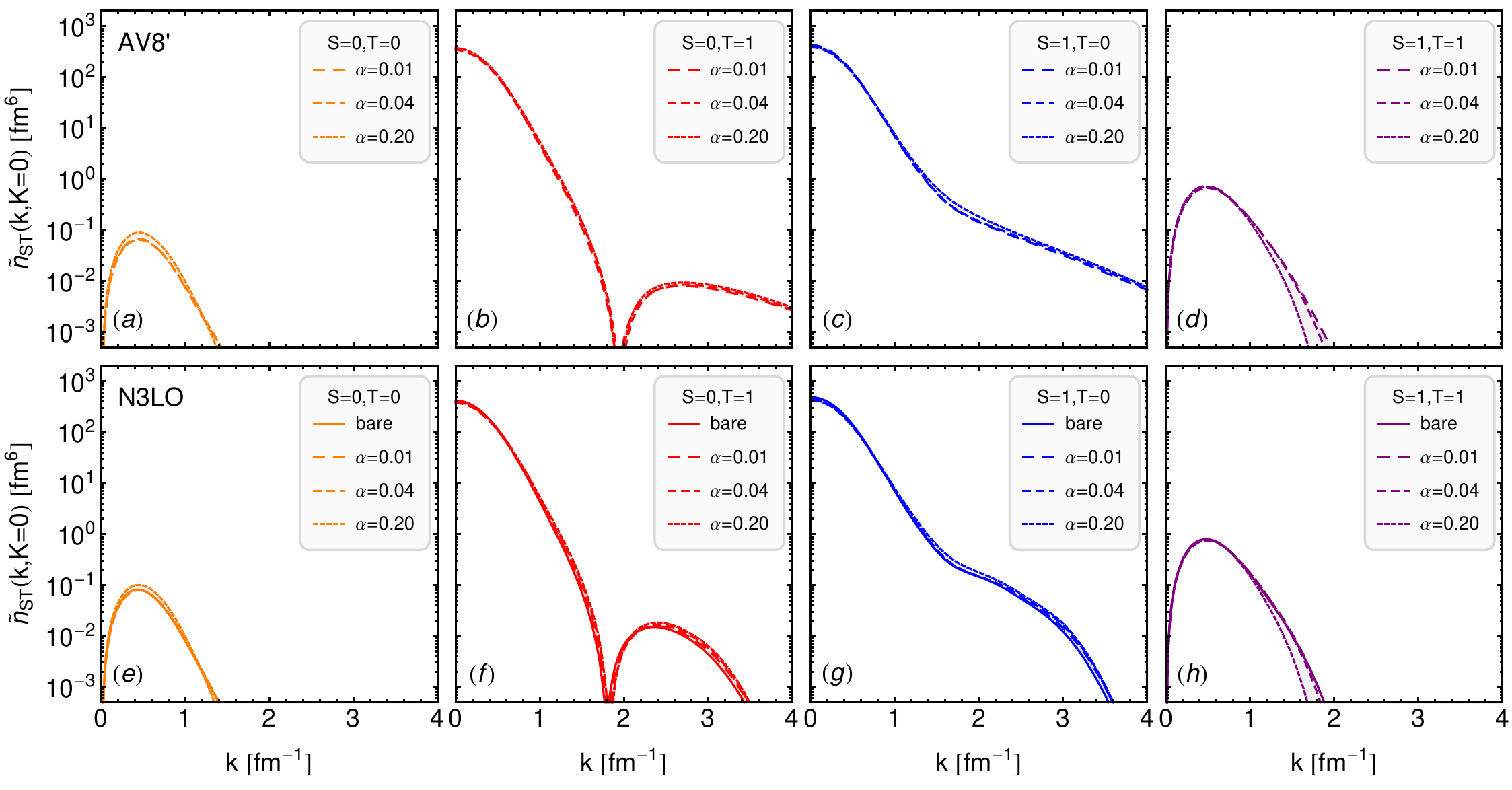}
  \caption{(Color online) AV8' (top) and N3LO (bottom) relative momentum distributions for pairs with vanishing
  pair momentum $K=0$ in the different spin-isospin channels obtained with SRG transformed
  density operators.} 
  \label{fig:cmtbdens-momentum-st}
\end{figure*}

In order to study this we
investigate the more exclusive joint probability
$\tilde{n}_{\alpha,ST}(k,K)$ to find a nucleon pair with spin $S$ and isospin
$T$ at relative momentum $k$ and total pair momentum $K$. It is calculated with
the two-body operator 
\begin{multline}
  \op{n}_{lST}(k,K) = \\
  \sum_{m M_S M_T} \ket{k l m, SM_S, TM_T} \bra{k l m, SM_S, TM_T} \\
  \otimes \sum_{LM} \ket{KLM} \left( \frac{A}{A-2} \right)^{3/2} \bra{KLM} \: ,
\end{multline}
where $\ket{KLM}$ denotes the spherical momentum representation of the relative momentum
of the pair with respect to the $A-2$ remaining nucleons. The factor 
$\left( \frac{A}{A-2} \right)^{3/2}$ originates from the transformation from Jacobi coordinates
to the coordinates $\vec{k}$ and $\vec{K}$, see Sec.~\ref{app:two-body-density-intrinsic}.
In this paper we consider only pair momentum $K=0$ for which the sum over $L,M$ reduces to $L=0, M=0$. 

The momentum distributions $\tilde{n}_{\alpha,ST}(k,K=0)$, obtained
by summing $\tilde{n}_{\alpha,lST}(k,K=0)$ over $l$, are displayed in
Fig.~\ref{fig:cmtbdens-momentum-st}.
One sees that in all $S,T$ channels the results are essentially independent on $\alpha$.
We can also observe that with increasing flow parameter $\alpha$ the momentum distributions 
$\tilde{n}_{\alpha,ST}(k)$, as shown in Fig.~\ref{fig:tbdens-momentum-st}, become more and more
similar to the momentum distributions $\tilde{n}_{\alpha,ST}(k,K=0)$ . 
This consistently tells us that for pairs with total momentum $K=0$ 
many-body correlations are not very important, as anticipated in the discussion above. 
The $K=0$ pairs are therefore the best candidates for experimental studies of short-range two-body
correlations. Similar considerations can be found in Refs.~\cite{wiringa08,alvioli13,arrington12}.

\begin{figure}
  \centering
  \includegraphics[width=\columnwidth]{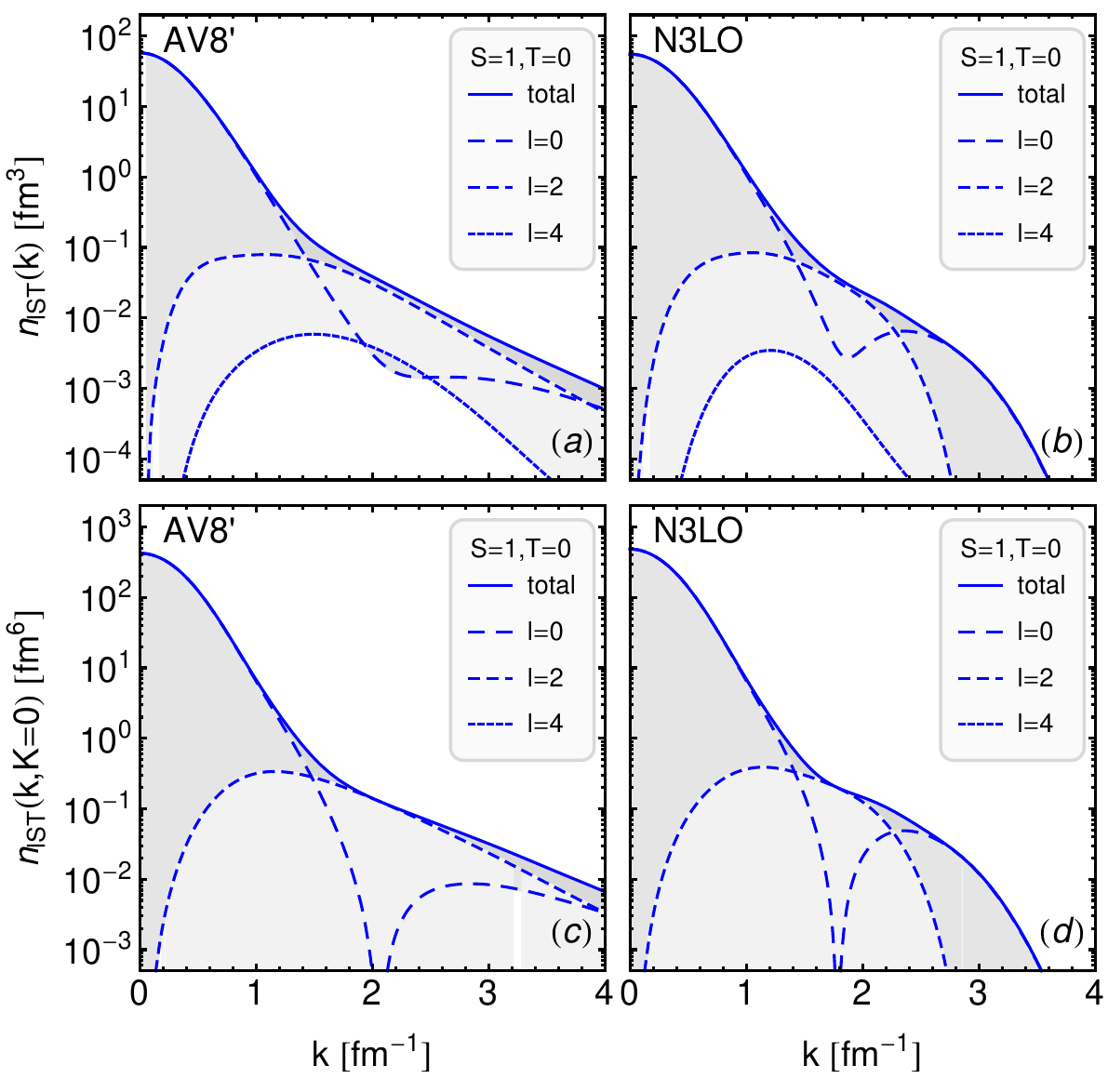}
  \caption{(Color online) Momentum distributions for the bare AV8' and N3LO interactions in the $S,T=1,0$ channel
  decomposed into contributions from pairs with relative orbital angular momentum $l=0,2,4$. For all
  pairs (top) and for pairs with vanishing pair momentum (bottom).}
  \label{fig:tbdens-momentum-l}
	\vspace{1ex}
  \includegraphics[width=\columnwidth]{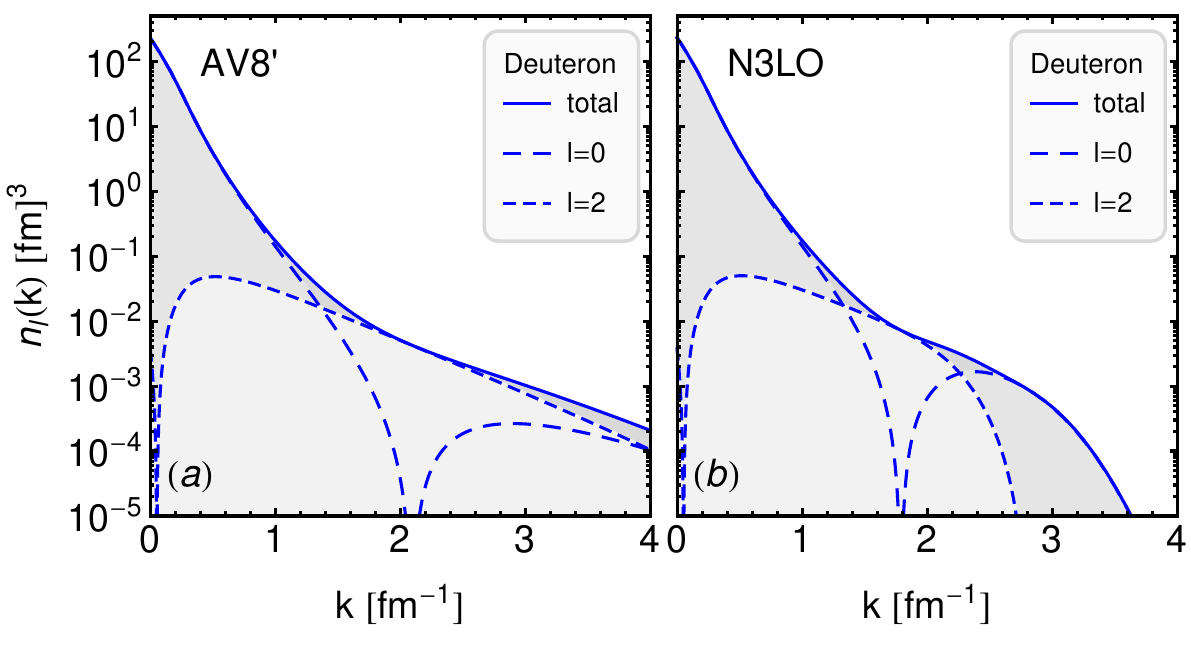}
  \caption{(Color online) Total, $l=0$ and $l=2$ momentum distributions in the deuteron ($S,T=1,0$) for 
 	the AV8' and the N3LO interactions.}
  \label{fig:tbdens-momentum-deuteron-l}
\end{figure}

We can also notice significant differences between the two even channels. In the $S,T=0,1$ channel the
momentum distribution has a node at relative momenta of about $1.8\,\fm^{-1}$. This is very
different in the $S,T=1,0$ channel. Here the momentum distribution does not show a minimum and
the number of pairs for relative momenta above about $1.5\,\fm^{-1}$ is significantly larger. This
difference is due to short-range tensor correlations that only contribute in the $S,T=1,0$ channel.
To illustrate this we show in Fig.~\ref{fig:tbdens-momentum-l} the
momentum distributions in the $S,T=1,0$ channel decomposed into their contributions 
from different $l$ for the AV8' and N3LO interactions. 
For the $K=0$ distributions we do not have the results for the bare AV8' interaction and 
used the SRG transformed densities $\tilde{n}^\mathrm{rel}_{lST}(k,K=0)$ for $\alpha = 0.01\,\fm^{-1}$
instead. Because of the very weak $\alpha$ dependence this should be equivalent to the exact result for the bare AV8' interaction. 
The distributions $\tilde{n}_{\alpha,lST}(k,K=0)$ in this channel are of particular
interest, because they show the dominant contribution from pairs with relative
angular momentum $l=2$ in the momentum region from $1.5$ to almost $4\,\fm^{-1}$ for
the AV8' and from $1.5$ to about $2.5\,\fm^{-1}$ for the N3LO interaction. These
$D$-wave pairs are directly reflecting the correlations induced by the tensor
force. In the AV8' case the tensor correlations are present at all
relative momenta and dominate over the $l=0$ contribution for momenta above $k\approx
1.5\,\fm^{-1}$. For N3LO the regularization cuts them off at higher momenta so that
they dominate only between $k \approx 1.5\,\fm^{-1}$ and $k \approx 2.5\,\fm^{-1}$. 
It is important to note that the tensor correlations contribute 
also at low momenta. There is no natural scale in the tensor correlations that would
allow to separate low and high momentum regions.

It is also interesting to compare the \nuc{4}{He} momentum distributions in the $S,T=1,0$
channel with those of the deuteron shown in Fig.~\ref{fig:tbdens-momentum-deuteron-l}. 
Whereas the momentum distributions in the low momentum region up to about $1.5\,\fm^{-1}$ 
are noticeably different, reflecting the differences in the long-range parts of the wave functions for
the loosely bound deuteron and the strongly bound \nuc{4}{He}, the momentum distributions are almost
indistinguishable for momenta above $1.5\,\fm^{-1}$. This is consistent with the 
observed universality of short-range correlations discussed in Ref.~\cite{src11}.  

If we compare the momentum distributions for the $S,T=1,0$ pairs integrated over all pair momenta $\vec{K}$
as shown in Fig.~\ref{fig:tbdens-momentum-l} (a) and (b) with those for the corresponding $K=0$ pairs
we notice that the node for the $S$-wave pairs has vanished and that we find additional contributions from higher
relative orbital angular momenta ($G$-wave pairs). This is consistent with our expectation
that many-body correlations play a greater role for pairs with large pair momentum $\vec{K}$. This is further
confirmed by a strong $\alpha$ dependence for these momentum distributions (not shown here). 
Alvioli \textit{et al.} \cite{alvioli13} found that the AV8' momentum distributions factorize in a relative
and a pair distribution for pair momenta $\vec{K}$ up to about $1\,\fm^{-1}$, whereas for higher pair momenta
the momentum distributions $n_{ST}(\vec{k},\vec{K})$ no longer factorize and depend on the relative
orientation of $\vec{k}$ and $\vec{K}$.

\subsection{Relative probabilities for $S,T$ pairs}

\begin{figure}
  \centering
  \includegraphics[width=\columnwidth]{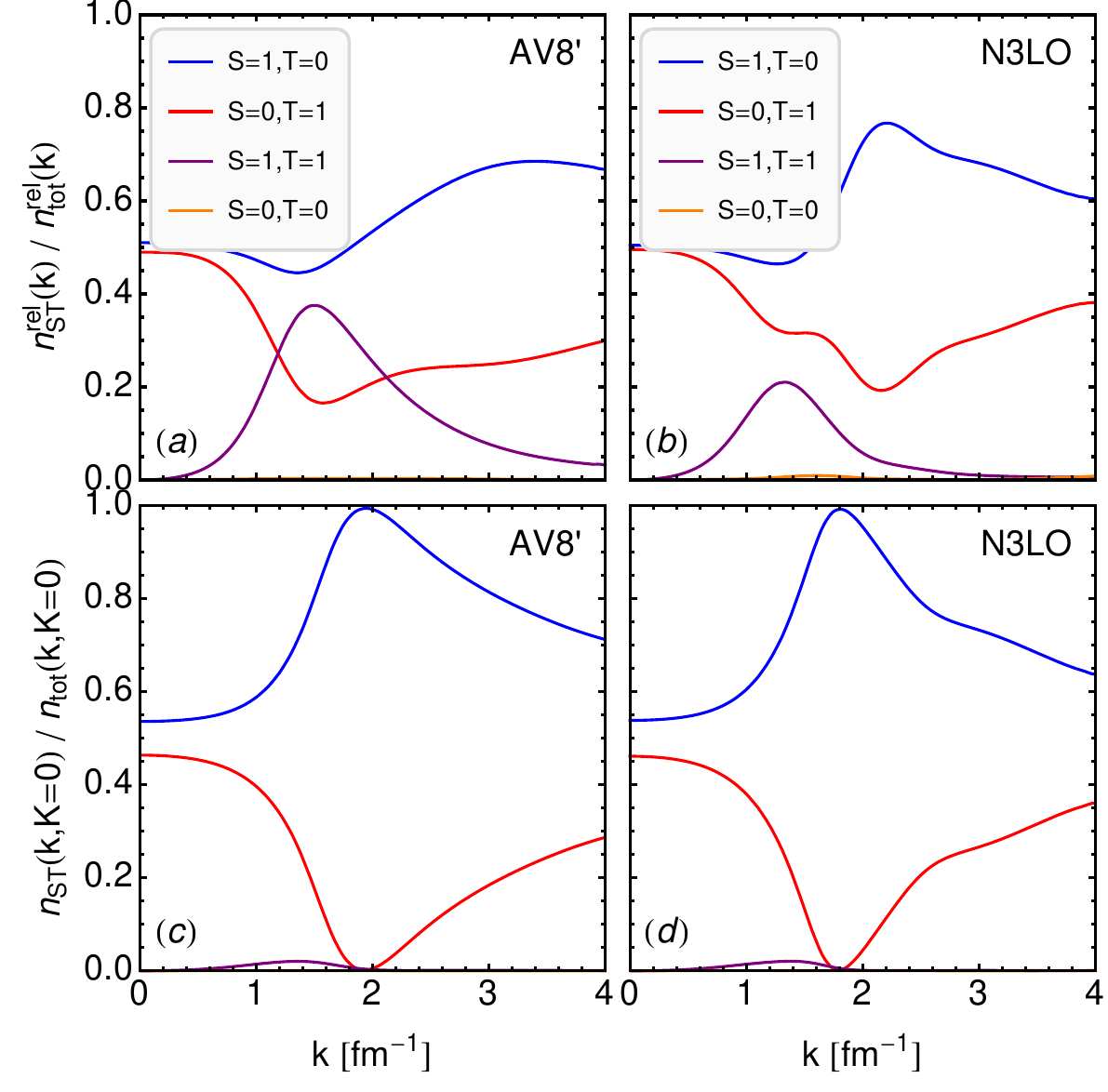}
  \caption{(Color online) Relative probability to find pairs in the different spin-isospin
  channels as a function of relative momentum. For all pairs (top) or only for pairs with
  pair momentum $K=0$ (bottom).}
  \label{fig:ST-probability}
\end{figure}

The strength of two- and many-body correlations strongly depends on the relative momenta of the pairs. To highlight
the relative importance of these correlations it is advantageous to look at the relative probabilities
\begin{equation}
	\frac{n^\mathrm{rel}_{ST}(k)}{\sum_{ST} n^\mathrm{rel}_{ST}(k)} \: , \quad
	\frac{n_{ST}(k,K=0)}{\sum_{ST} n_{ST}(k,K=0)}
\end{equation}
to find pairs in a given $S,T$ channel as a function of relative momentum $k$. In Fig.~\ref{fig:ST-probability} we 
show these relative probabilities for the bare AV8' and N3LO interactions.
If we look at the relative probabilities for all pairs with relative momentum $k$ it is obvious that the $S,T=1,1$ pairs
contribute significantly, especially around $k \approx 1.5\,\fm^{-1}$. The total number of $S,T=1,1$ pairs is much smaller
than the number of pairs in the even channels as shown in Tab.~\ref{tab:pairs}, but in this mid-momentum region
the number of pairs in the $S,T=1,1$ channel
is comparable to those in the even channels. This is the region where many-body correlations have the largest
effect. For small relative momenta the relative probabilities are dominated by the mean-field, for relative
momenta above about $3\,\fm^{-1}$ in case of the AV8' and about $2.5\,\fm^{-1}$ in case of the N3LO
interaction the relative probabilities are dominated by short-range correlations. This influence of
many-body correlations is also related to a strong dependence on $\alpha$. With increasing flow parameter
$\alpha$ the relative probabilities for all pairs become more and more similar to the relative
probabilities of the $K=0$ pairs shown in Fig.~\ref{fig:ST-probability} (c) and (d). As there is no significant
$\alpha$ dependence for the $K=0$ momentum distributions, the relative probabilities for the $K=0$ pairs are
independent from $\alpha$ as well and are therefore not sensitive to many-body correlations.

It is interesting to note that the relative probabilities for $K=0$ the pairs are quite similar for the 
AV8' and N3LO interactions, even for very large relative momenta, whereas the absolute values of the 
momentum distributions are very different. Differences in the relative probabilities
between AV8' and N3LO reflect differences in the relative importance of tensor correlations for the
two interactions due to differences in the regularization of the tensor force.

\subsection{Relative probabilities for $pn$ and $pp$ pairs} 

In an experiment one measures protons and neutrons and not $S,T$ pairs.
Therefore we define the operator in two-body space that measures the probability to find a
pair of two protons ($pp$:~$M_T=1$) or a proton-neutron pair
($pn+np$:~$M_T=0$) at relative momentum $k$ and pair momentum $K$ as
\begin{multline}
  \op{n}_{M_T}(k,K) = \\
  \sum_{l m, S M_S, T} \ket{k l m, SM_S, TM_T} \bra{k l m, SM_S, TM_T} \\
  \otimes \sum_{LM} \ket{KLM} \left( \frac{A}{A-2} \right)^{3/2} \bra{KLM} \: .
\end{multline}

In the case of \nuc{4}{He} one $T=0$ pair corresponds to one $pn$-pair, and
one $T=1$ pair to $\tfrac{1}{3}$ of a $pp$, $\tfrac{1}{3}$ of a $nn$ and $\tfrac{1}{3}$ of a $pn$ pair.

\begin{figure}
  \centering
  \includegraphics[width=\columnwidth]{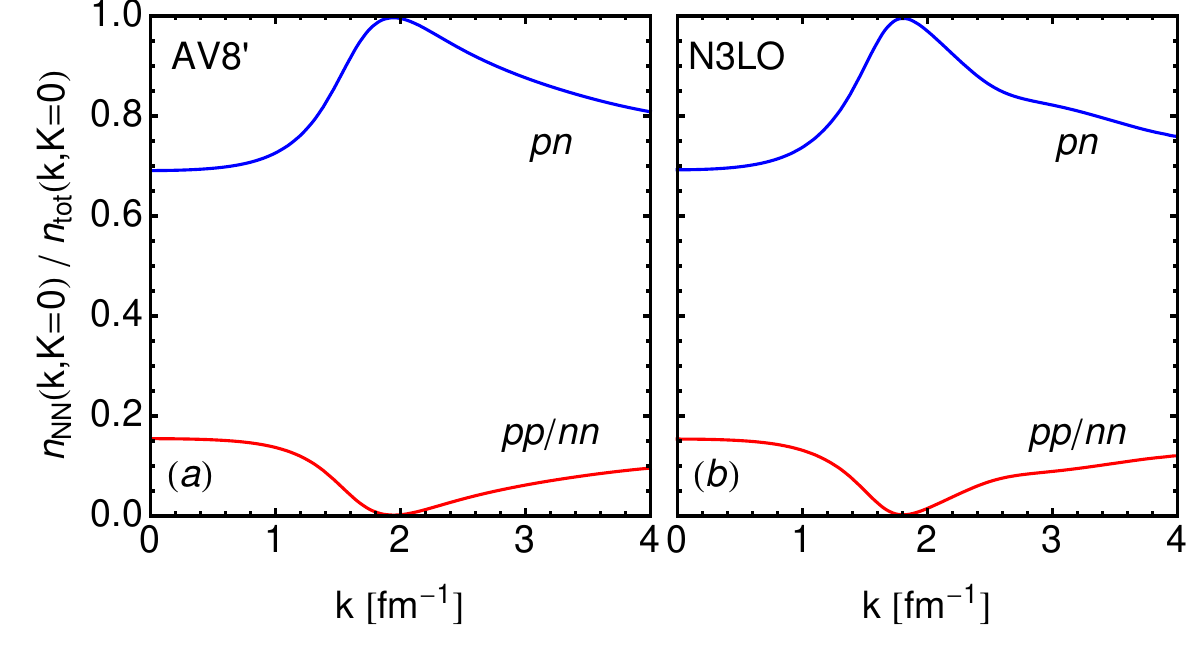}
  \caption{(Color online) Relative probability to find $pn$ or $pp$ pairs
	with pair momentum $K=0$ as a function of relative momentum.}
  \label{fig:pn-pp-probability}
\end{figure}

The corresponding relative probabilities for $K=0$ pairs
\begin{equation}
	\frac{n_{M_T}(k, K=0)}{\sum_{M_T} n_{M_T}(k, K=0)}
\end{equation}
are shown in Fig.~\ref{fig:pn-pp-probability}. The first observation is that the relative probabilities
are rather similar for the AV8' and N3LO interactions. At low momenta both show a ratio close to 
$\tfrac{1}{4}$ to find $pp$ versus $pn$ pairs. This is to be expected because
an uncorrelated system of two protons and two neutrons can form one $pp$-pair, one $nn$-pair
but four $pn$-pairs.

Around $k \approx 1.8\,\fm^{-1}$ the minimum in the $S,T = 0,1$ channel (see
Fig.~\ref{fig:cmtbdens-momentum-st}) together with the $l=2$ contribution from
the tensor interaction in the $S,T = 1,0$ channel (see Fig.~\ref{fig:tbdens-momentum-l}) enhances the relative
probability to find a $pn$ pair to almost 100\%. This dominance of $pn$ pairs has
been observed in exclusive two-nucleon knockout experiments \cite{tang03,subedi08}.
Recently the $pn$ to $pp$ ratio has been measured for relative momenta $k$ from $2.5\,\fm^{-1}$
up to almost $4\,\fm^{-1}$ \cite{korover14} showing an increase in the $pp$/$pn$ ratio. Within
the experimental uncertainties the data agree with our results for both AV8' and N3LO
interactions.

\section{Summary and Conclusions}
\label{sec:summary}

In this paper we applied the SRG formalism for the calculation of relative density and
momentum distributions of \nuc{4}{He}. The \nuc{4}{He} ground state wave functions are calculated in
the NCSM with the SRG evolved AV8' and N3LO interactions in two-body approximation. 
Two-body densities in coordinate and momentum space calculated with the unevolved 
density operators illustrate how short-range correlations are eliminated by the SRG evolution.
With increasing flow parameter $\alpha$ the interaction gets `softer' and the wave 
functions become essentially uncorrelated mean-field wave functions without correlation
holes and high-momentum components. The short-range or high-momentum information can be recovered by
calculating two-body densities with the SRG evolved density operators, again in two-body
approximation. Using these effective density operators we see a dependence of the
two-body densities on the flow parameter $\alpha$. This $\alpha$ dependence is due to
missing contributions from three- and four-body terms in the effective operators,
that are omitted in the two-body approximation. We find that not all components of the
two-body density are equally affected by many-body correlations. The momentum distributions
for pairs with pair momentum $K=0$ show only a very weak $\alpha$ dependence and
therefore provide direct access to two-body short-range correlations. The momentum
distributions integrated over all pair momenta on the other hand have a sizeable
$\alpha$ dependence. This $\alpha$ dependence is particularly strong in the $S,T=0,1$
and $S,T=1,1$ channels for momenta around $1.5\,\fm^{-1}$. We identified three-body
correlations induced by the strong tensor force as the
main contributor to these many-body correlations.

Our results for the AV8' momentum distributions agree with previous results \cite{schiavilla07,wiringa08,wiringa14,alvioli13} 
using variational Monte-Carlo and few-body approaches. We confirm the important role of tensor
correlations that explain the experimentally observed dominance of $pn$ over $pp$ pairs
in exclusive two-nucleon knock-out with large momentum transfer. We also show that the
chiral N3LO interaction provides very similar results for the momentum distributions
at least up to relative momenta of $2.5-3.0\,\fm^{-1}$. This includes the role of tensor
correlations that are very similar for both interactions. At larger relative momenta the
N3LO momentum distributions fall off much faster than the AV8' momentum distributions
as would be expected from the relatively soft cutoff employed in the regularization
of the N3LO interaction. These differences are however mostly hidden in the ratios of $pn$
versus $pp$ pairs as a function of relative momenta as investigated in two-nucleon
knockout experiments. Recently new chiral interactions
with different regularization schemes have been proposed \cite{wendt14,epelbaum14}.
It will be interesting to see how these affect the short-range behavior and momentum
distributions.

In this paper we investigated the \nuc{4}{He} nucleus as it allowed us to compare SRG evolved with bare
interactions. SRG evolved soft Hamiltonians however will allow us to study heavier nuclei 
in the $p$-shell using the NCSM. One interesting question is the isospin dependence of the
short-range correlations \cite{sargsian14,hen14} in asymmetric nuclei. Neutron halos or skins might
be a useful laboratory for the study of neutron matter. The present study also did not include
three-body forces. Wiringa \textit{et al.} \cite{wiringa14} find only small differences in the proton
momentum distributions obtained with AV18 alone and AV18 together with an UX three-body
interaction. We would expect that three-body interactions would not significantly change
the short-range two-body correlations in the $K=0$ pairs, they may however have a significant
effect on the many-body correlations and therefore change the results for the momentum
distributions integrated over all pair momenta.

\appendix

\section{Talmi-Moshinsky transformation}
\label{app:talmi}

For an orthogonal transformation of the coordinates ($\vec{x}_1$,
$\vec{x}_2$) into ($\vec{X}$, $\vec{x}$) with the mass ratio
$d=A_1/A_2$
\begin{equation}
	\label{eq:talmi-coord}
  \begin{pmatrix} \vec{X} \\ \vec{x} \end{pmatrix} =
  \begin{pmatrix}
    \sqrt{\frac{d}{1+d}} & \sqrt{\frac{1}{1+d}} \\
    \sqrt{\frac{1}{1+d}} & -\sqrt{\frac{d}{1+d}}
  \end{pmatrix}
  \begin{pmatrix} \vec{x}_1 \\ \vec{x}_2 \end{pmatrix}
\end{equation}
the product of harmonic oscillator wave functions in coordinates
$\vec{x}_1$ and $\vec{x}_2$ can be expressed with the Talmi-Moshinsky
brackets $\talmi{NL}{nl}{n_1 l_1}{n_2 l_2}{\Lambda}{d}$ in the new coordinates as:
\begin{multline}
	\label{eq:talmi}
  \sum_{m_1 m_2} \cg{l_1}{m_1}{l_2}{m_2}{\Lambda}{\lambda} 
  \varphi^b_{n_1 l_1 m_1}(\vec{x}_1)  \varphi^b_{n_2 l_2 m_2}(\vec{x}_2) = \\
  \sum_{NL,nl} \talmi{NL}{nl}{n_1 l_1}{n_2 l_2}{\Lambda}{d} \\
  \times \sum_{M m} \cg{L}{M}{l}{m}{\Lambda}{\lambda} 
  \varphi^b_{NLM}(\vec{X})  \varphi^b_{nlm}(\vec{x})
\end{multline}

All harmonic oscillator wave functions have the same oscillator
parameter $b$.  See Ref.~\cite{kamuntavicius01} for further
properties.

\section{Jacobi coordinates}
\label{app:jacobi-coordinates}

The wave function and two-body densities in the intrinsic system will be expressed in
Jacobi coordinates. We follow \cite{navratil04} and use mass-scaled Jacobi coordinates 
that fulfill the orthogonality condition \eqref{eq:talmi-coord}:
\begin{align}
  \vec{\xi}_0 & = \sqrt{\frac{1}{A}} \left[ \vec{x}_1 + \ldots + \vec{x}_A \right] \\
  \vec{\xi}_1 & = \sqrt{\frac{1}{2}} \left[ \vec{x}_1 - \vec{x}_2 \right] \\
  \vec{\xi}_2 & = \sqrt{\frac{2}{3}} \left[ \frac{1}{2} ( \vec{x}_1 + \vec{x}_2 ) - \vec{x}_3 \right] \\
  & \vdots \notag \\
  \vec{\xi}_{A-1} & = \sqrt{\frac{A-1}{A}} \left[ \frac{1}{A-1} ( \vec{x}_1 + \ldots + \vec{x}_{A-1} ) - \vec{x}_A \right]
\end{align}
The transformation from the coordinates $(\vec{x}_1, \ldots, \vec{x}_A)$ to $(\vec{\xi}_0, \ldots, \vec{\xi}_{A-1})$
is orthogonal. Translational invariant one-body densities can be expressed in these Jacobi coordinates. For the 
two-body densities we prefer however a different set of Jacobi coordinates where we have the distance between the
nucleons and the distance of the center of mass of the pair with respect to the rest of the nucleus as coordinates. 
This can be achieved by an additional orthogonal transformation from the coordinates $(\vec{\xi}_{A-2}, \vec{\xi}_{A-1})$ 
into the coordinates
$(\vec{\eta}, \vec{\vartheta})$
\begin{multline}
  \vec{\eta} = \sqrt{\frac{2(A-2)}{A}} \\
  \times \Bigl[ \frac{1}{A-2} ( \vec{x}_1 + \ldots + \vec{x}_{A-2}) -
  \frac{1}{2} (\vec{x}_{A-1} + \vec{x}_{A}) \Bigr] \: ,
\end{multline}
\begin{equation}
  \vec{\vartheta} = \sqrt{\frac{1}{2}} \bigl[ \vec{x}_{A-1} - \vec{x}_A \bigr] \: .
\end{equation}

\section{Two-body densities in the laboratory system}
\label{app:two-body-density-lab}

In the NCSM the wave function is expanded in a harmonic oscillator 
single-particle basis with oscillator parameter $b$
\begin{equation}
  \ket{q} = \ket{b; nlm, m_s, m_t} \: .
\end{equation}

The NCSM uses second quantization techniques and the two-body density
in the harmonic oscillator basis can be expressed as
\begin{equation}
	\label{eq:two-body-den-sp}
  \rho^\mathrm{lab}_{q_1' q_2'; q_1 q_2} = 
  \matrixe{\Psi}{\adj{\op{a}}_{q_1} \adj{\op{a}}_{q_2} \op{a}_{q_2'} \op{a}_{q_1'}}{\Psi} \: .
\end{equation}

For the discussion of two-body correlations it is natural to transform from
single-particle to pair coordinates. In the laboratory system we define relative 
and pair coordinates as
\begin{equation}
 \vec{r} = \vec{x}_1-\vec{x}_2, \quad \vec{R} = \frac{1}{2} (\vec{x}_1 + \vec{x}_2) \: ,
\end{equation}
and the conjugate coordinates in momentum space as
\begin{equation}
 \vec{k} = \frac{1}{2} (\vec{p}_1-\vec{p}_2), \quad \vec{K} = \vec{p}_1 + \vec{p}_2 \: .
\end{equation}

These coordinates differ from the ones in \eqref{eq:talmi-coord} by factors of 
$\sqrt{2}$ and $1/\sqrt{2}$ so that the oscillator parameters for the relative and
the pair motion have to changed to $b_\mathrm{rel} = \sqrt{2}\, b$ and 
$b_\mathrm{pair} = \frac{1}{\sqrt{2}}\, b$ respectively. Using \eqref{eq:talmi} and
coupling orbital angular momenta, spins and isospins the two-body density \eqref{eq:two-body-den-sp} can be
transformed into $\rho^\mathrm{lab}_{qQ;q'Q'}$.

Using this new basis
\begin{equation}
  \ket{qQ} = \ket{b_\mathrm{rel}; nlm SM_S TM_T} \otimes \ket{b_\mathrm{pair}; NLM}
\end{equation}
we can define the density operator in two-body space as
\begin{equation}
	\label{eq:two-body-den-op}
  \op{R}^\mathrm{lab} = \sum_{qQ;q'Q'} \ket{qQ} \rho^\mathrm{lab}_{qQ;q'Q'} \bra{q'Q'} \: .
\end{equation}
$\op{R}^\mathrm{lab}$ allows to easily express the two-body density in coordinate space as a function of 
relative distance $\vec{r}$ and pair position $\vec{R}$
\begin{equation}
  \rho^\mathrm{lab}(\vec{r},\vec{R}) = 
  \Tr_2 \left( \op{R}^\mathrm{lab} \, \ket{\vec{r};\vec{R}} \bra{\vec{r};\vec{R}} \right) \; ,
\end{equation}
or as function of relative momentum $\vec{k}$ and pair momentum $\vec{K}$
\begin{equation}
  n^\mathrm{lab}(\vec{k},\vec{K}) = \Tr_2 \left( \op{R}^\mathrm{lab} \, \ket{\vec{k};\vec{K}} \bra{\vec{k};\vec{K}} \right) \: .
\end{equation}
This can be trivially extended to calculate off-diagonal densities or densities for pairs of a given
spin and isospin.

These densities can also be expressed as
\begin{multline}
	\rho^\mathrm{lab}(\vec{r},\vec{R}) = \\
	\sum_{i<j} \matrixe{\Psi}{\delta^3(\op{\vec{x}}_i-\op{\vec{x}}_j - \vec{r}) \delta^3(\frac{1}{2}(\op{\vec{x}}_i+\op{\vec{x}_j}) - \vec{R})}{\Psi}
\end{multline}
and can be obtained from the wave functions by integrating out $A-2$ coordinates 
(we have omitted spin- and isospin indices for brevity)
\begin{multline}
  \rho^\mathrm{lab}(\vec{r},\vec{R}) = \frac{A(A-1)}{2} \int d^3x_1 \cdots d^3x_{A-2} \\
  \conj{\Psi(\vec{x}_1, \ldots, \vec{x}_{A-2}, \vec{R}+\tfrac{1}{2}\vec{r}, \vec{R}-\tfrac{1}{2}\vec{r})}\\
   \times \Psi(\vec{x}_1, \ldots, \vec{x}_{A-2}, \vec{R}+\tfrac{1}{2}\vec{r}, \vec{R}-\tfrac{1}{2}\vec{r}) \: ,
\end{multline}
where we have used the antisymmetry of the wave function.

\section{Two-body densities in the intrinsic system}
\label{app:two-body-density-intrinsic}

The two-body density $\rho^\mathrm{lab}_{qQ;q'Q'}$ is calculated in the harmonic oscillator
basis of the NCSM localized at the origin of the coordinate system and is therefore not
translationally invariant. However the wave function in the NCSM factorizes into an intrinsic
wave function only depending on relative coordinates and the total center-of-mass wave 
function in the ground state of the harmonic oscillator \cite{navratil00}
\begin{equation} \label{eq:ncsm-factorization}
  \Psi(\vec{x}_1, \ldots, \vec{x}_A) = 
  \Psi_\mathrm{int}(\vec{\xi}_1, \ldots, \vec{\xi}_{A-3}, \vec{\eta}, \vec{\vartheta})
  \varphi^b_{000}(\vec{\xi}_0) \: .
\end{equation}

This allows us to relate the two-body density in the laboratory system $\rho^\mathrm{lab}_{qQ;q'Q'}$
with the two-body density in the intrinsic system. The derivation is lengthy but straightforward and follows
the derivation of the translational invariant one-body density in Ref.~\cite{navratil04}. 

One starts with the two-body density in the laboratory system as given from the wave functions written in the
coordinates $\vec{x}_1, \ldots, \vec{x}_{A-2}, \vec{X}_2 = \frac{1}{\sqrt{2}} (\vec{x}_{A-1}+\vec{x}_A)$ and 
$\vec{\vartheta} = \frac{1}{\sqrt{2}} (\vec{x}_{A-1} - \vec{x}_A)$:
\begin{widetext}
\begin{multline} \label{eq:density-lab}
  \rho^\mathrm{lab}_{q', N_2' L_2' M_2'; q N_2 L_2 M_2} =
  \frac{A(A-1)}{2} \int d^3x_1 \cdots d^3x_{A-2} \, d^3X_2 \, d^3\vartheta 
  \int d^3x_1' \cdots d^3x_{A-2}' \ d^3X_2' \, d^3\vartheta' \\
  \times \conj{\Psi(\vec{x}_1, \ldots, \vec{x}_{A-2}, \vec{X}_2, \vec{\vartheta})}
  \delta^3(\vec{x}_1-\vec{x}_1') \cdots \delta^3(\vec{x}_{A-2}-\vec{x}_{A-2}')
  \Psi(\vec{x}_1', \ldots, \vec{x}_{A-2}', \vec{X}_2', \vec{\vartheta}') \\
  \times \conj{\varphi_{N_2' L_2' M_2'}(\vec{X}_2')} \, \varphi_{N_2 L_2 M_2}(\vec{X}_2) \:
  \conj{\varphi_{q'}(\vec{\vartheta}')} \, \varphi_q(\vec{\vartheta}) 
\end{multline}

We now perform an orthogonal coordinate transformation from $(\vec{x}_1, \ldots, \vec{x}_{A-2}, \vec{X}_2)$ to
$(\vec{X}_{A-2}, \vec{\xi}_1, \ldots, \vec{\xi}_{A-3})$ and correspondingly for the primed coordinates. Next one uses
the properties of the delta function to obtain delta functions in the Jacobi coordinates $\vec{\xi}_k$ and in $\vec{X}_{A-2} = \frac{1}{\sqrt{A-2}} \sum_{i=1}^{A-2} \vec{x}_i$.  
The delta function in $\vec{X}_{A-2}$ is expanded in the harmonic oscillator basis
\begin{equation}
 \delta^3(\vec{X}_{A-2}-\vec{X}_{A-2}') = \sum_{N_{A-2} L_{A-2} M_{A-2}}
 \varphi_{N_{A-2} L_{A-2} M_{A-2}}(\vec{X}_{A-2}) \,
 \conj{\varphi_{N_{A-2} L_{A-2} M_{A-2}}(\vec{X}_{A-2}')} \: .
\end{equation}

In the next step a second orthogonal coordinate transformation from $(\vec{X}_{A-2}, \vec{X}_2)$ to 
$(\vec{\xi}_0, \vec{\eta})$ is performed employing the Talmi-Moshinsky transformation for the harmonic
oscillator wave functions, rewriting products 
$\varphi_{N_{A-2} L_{A-2} M_{A-2}}(\vec{X}_{A-2}) \varphi_{N_2 L_2 M_2}(\vec{X}_2)$ with linear
combinations of products
$\varphi_{N_\mathrm{cm} L_\mathrm{cm} M_\mathrm{cm}}(\vec{\xi}_0) \varphi_{N L M}(\vec{\eta})$.

If one uses \eqref{eq:ncsm-factorization} one can now integrate over $\vec{\xi}_0$ and $\vec{\xi}_0'$ to 
express the density matrix in the laboratory system with the intrinsic wave functions so that the density
matrix in the laboratory system can be related to the density matrix in the intrinsic system:
\begin{multline} \label{eq:density-intrinsic}
  \rho_{q', N' L' M'; q N L M} =
  \frac{A(A-1)}{2} \int d^3\xi_1 \cdots d^3\xi_{A-3} \, d^3\eta \, d^3\eta' \, d^3\vartheta \, d^3\vartheta' \\
  \times \conj{\Psi_\mathrm{int}(\vec{\xi}_1, \ldots, \vec{\xi}_{A-3}, \vec{\eta}, \vec{\vartheta})}
  \Psi_\mathrm{int}(\vec{\xi}_1, \ldots, \vec{\xi}_{A-3}, \vec{\eta}', \vec{\vartheta}') \:
 	\conj{\varphi_{N' L' M'}(\vec{\eta}')} \, \varphi_{N L M}(\vec{\eta}) \:
  \conj{\varphi_{q'}(\vec{\vartheta}')} \, \varphi_q(\vec{\vartheta})
\end{multline}

In this paper we only discuss scalar densities in the pair coordinates (integrating over all pair momenta $\vec{K}$ or $K=0$). 
Then $L=L'$ and $M=M'$ and we can use the completeness relations for the Clebsch-Gordan coefficients to
obtain the relation between the density matrices in the laboratory \eqref{eq:density-lab} and the intrinsic system \eqref{eq:density-intrinsic}
\begin{equation}
 \sum_{M_2} \rho^\mathrm{lab}_{q' N_2' L_2 M_2; q N_2 L_2 M_2} = 
 \sum_{NN'L} M^{(0)}_{N_2' L_2,N_2 L_2;N'L,NL}
 \sum_{M} \rho_{q' N' L M; q N L M} \: .
 \label{eq:tbdens-ti}
\end{equation}
with the transformation matrix $M^{(0)}$
\begin{equation}
  M^{(0)}_{N_2' L_2, N_2 L_2; N'L, NL} =
  \sum_{N_{A-2} L_{A-2}}
  \talmi{00}{NL}{N_{A-2} L_{A-2}}{N_2 L_2}{L}{(A-2)/2}
  \talmi{00}{N'L}{N_{A-2} L_{A-2}}{N_2' L_2}{L}{(A-2)/2}
\end{equation}
given by the Talmi-Moshinsky brackets with the mass ratio $d=\frac{A-2}{2}$.
This reflects the distribution of the oscillator quanta among the 
nucleon pair and the $A-2$ remaining nucleons. The translationally invariant
two-body density averaged over the orientations of the pair momenta
$\sum_{M} \rho_{q' N' L M; q N L M}$ is then obtained from the two-body density matrix in the
laboratory system by inverting Eq.~\eqref{eq:tbdens-ti}.

The two-body density in the Jacobi coordinates $\vec{\eta}$ and $\vec{\vartheta}$ is given by
\begin{equation}
  \rho^\mathrm{jac}(\vec{\vartheta}, \vec{\eta}) = 
  \sum_{q NLM, q' N'L'M'} \rho_{q' N' L' M'; q N L M} \:
  \conj{\varphi^b_{NLM}(\vec{\eta})} \, \varphi^b_{N'L'M'}(\vec{\eta}) \:
  \conj{\varphi^b_{q}(\vec{\vartheta})} \, \varphi^b_{q'}(\vec{\vartheta})
\end{equation}

The mass-scaled Jacobi coordinates have technical advantages, however we prefer to express the
two-body densities in the more intuitive coordinates
\begin{equation}
  \vec{r} = \vec{x}_{A-1} - \vec{x}_A, \quad 
  \vec{R} = \frac{1}{2} (\vec{x}_{A-1} + \vec{x}_{A}) - \frac{1}{A-2} ( \vec{x}_1 + \ldots + \vec{x}_{A-2})
\end{equation}
that are related to the Jacobi coordinates by
\begin{equation} \label{eq:ti-pair-coord}
  \vec{r} = \sqrt{2} \, \vec{\vartheta}, \quad \vec{R} = - \sqrt{\frac{A}{2(A-2)}} \, \vec{\eta} \: .
\end{equation}
The conjugate variables to $\vec{r}$ and $\vec{R}$ are the relative momentum of the nucleons 
in the pair $\vec{k}$ and the relative momentum of the pair with respect to the rest of the 
nucleus $\vec{K}$ (in the center-of-mass system this is the same as the momentum of the pair).
\begin{equation}
  \vec{k} = \frac{1}{2} (\vec{p}_{A-1} - \vec{p}_A), \quad
  \vec{K} = \frac{A-2}{A} (\vec{p}_{A-1} + \vec{p}_A) - \frac{2}{A} ( \vec{p}_1 + \ldots + \vec{p}_{A-2})
\end{equation}
The two-body density in these coordinates is then
\begin{equation}
	\begin{split}
  \rho(\vec{r},\vec{R}) &=
  \left(\frac{A-2}{A}\right)^{3/2} 
  \rho^\mathrm{jac}\biggl(\frac{1}{\sqrt{2}} \vec{r}, -\sqrt{\frac{2(A-2)}{A}} \vec{R} \biggr) \\
  &= \left(\frac{A-2}{A}\right)^{3/2} \sum_{q NLM, q' N'L'M'} \rho_{q' N' L' M'; q N L M} \:
  \conj{\varphi^{b_\mathrm{pair}}_{NLM}(\vec{R})} \, \varphi^{b_\mathrm{pair}}_{N'L'M'}(\vec{R}) \:
  \conj{\varphi^{b_\mathrm{rel}}_{q}(\vec{r})} \, \varphi^{b_\mathrm{rel}}_{q'}(\vec{r})
  \end{split}
\end{equation}
Note that the oscillator parameter for the pair motion $b_\mathrm{pair} = \sqrt{\frac{A}{2(A-2)}}\, b$
is different than in the laboratory system, whereas the oscillator parameter $b_\mathrm{rel} = \sqrt{2}\, b$ is
the same. The factor $\left(\frac{A-2}{A}\right)^{3/2}$ is the Jacobian 
of the nonorthogonal transformation from the coordinates $\vec{\vartheta}$ and $\vec{\eta}$ to $\vec{r}$ and
$\vec{R}$ in \eqref{eq:ti-pair-coord}.

\end{widetext}

As in the laboratory system a convenient notation is obtained by defining the density operator in
two-body space
\begin{equation}
	\label{eq:two-body-den-op}
  \op{R} = \sum_{qQ;q'Q'} \ket{qQ} \rho_{qQ;q'Q'} \bra{q'Q'}
\end{equation}
with 
\begin{equation}
  \ket{qQ} = \ket{b_\mathrm{rel}; nlm SM_S TM_T} \otimes \ket{b_\mathrm{pair}; NLM} \: .
\end{equation}
The two-body densities in coordinate space can then be expressed as
\begin{equation}
  \rho(\vec{r},\vec{R}) = 
  \Tr_2 \left( \op{R} \> \ket{\vec{r};\vec{R}} \left(\frac{A-2}{A}\right)^{3/2} \bra{\vec{r};\vec{R}} \right)
\end{equation}
and in momentum space as
\begin{equation}
  n(\vec{k},\vec{K}) = 
  \Tr_2 \left( \op{R} \> \ket{\vec{k};\vec{K}} \left(\frac{A}{A-2}\right)^{3/2} \bra{\vec{k};\vec{K}} \right) \: .
\end{equation}

\acknowledgements
We acknowledge support by the ExtreMe Matter Institute EMMI in the framework
of the Helmholtz Alliance HA216/EMMI and by JSPS KAKENHI Grant Number 25800121.

\bibliography{src}

\begin{thebibliography}{51}%
\makeatletter
\providecommand \@ifxundefined [1]{%
 \@ifx{#1\undefined}
}%
\providecommand \@ifnum [1]{%
 \ifnum #1\expandafter \@firstoftwo
 \else \expandafter \@secondoftwo
 \fi
}%
\providecommand \@ifx [1]{%
 \ifx #1\expandafter \@firstoftwo
 \else \expandafter \@secondoftwo
 \fi
}%
\providecommand \natexlab [1]{#1}%
\providecommand \enquote  [1]{``#1''}%
\providecommand \bibnamefont  [1]{#1}%
\providecommand \bibfnamefont [1]{#1}%
\providecommand \citenamefont [1]{#1}%
\providecommand \href@noop [0]{\@secondoftwo}%
\providecommand \href [0]{\begingroup \@sanitize@url \@href}%
\providecommand \@href[1]{\@@startlink{#1}\@@href}%
\providecommand \@@href[1]{\endgroup#1\@@endlink}%
\providecommand \@sanitize@url [0]{\catcode `\\12\catcode `\$12\catcode
  `\&12\catcode `\#12\catcode `\^12\catcode `\_12\catcode `\%12\relax}%
\providecommand \@@startlink[1]{}%
\providecommand \@@endlink[0]{}%
\providecommand \url  [0]{\begingroup\@sanitize@url \@url }%
\providecommand \@url [1]{\endgroup\@href {#1}{\urlprefix }}%
\providecommand \urlprefix  [0]{URL }%
\providecommand \Eprint [0]{\href }%
\providecommand \doibase [0]{http://dx.doi.org/}%
\providecommand \selectlanguage [0]{\@gobble}%
\providecommand \bibinfo  [0]{\@secondoftwo}%
\providecommand \bibfield  [0]{\@secondoftwo}%
\providecommand \translation [1]{[#1]}%
\providecommand \BibitemOpen [0]{}%
\providecommand \bibitemStop [0]{}%
\providecommand \bibitemNoStop [0]{.\EOS\space}%
\providecommand \EOS [0]{\spacefactor3000\relax}%
\providecommand \BibitemShut  [1]{\csname bibitem#1\endcsname}%
\let\auto@bib@innerbib\@empty
\bibitem [{\citenamefont {Wiringa}\ \emph {et~al.}(1995)\citenamefont
  {Wiringa}, \citenamefont {Stoks},\ and\ \citenamefont
  {Schiavilla}}]{wiringa95}%
  \BibitemOpen
  \bibfield  {author} {\bibinfo {author} {\bibfnamefont {R.~B.}\ \bibnamefont
  {Wiringa}}, \bibinfo {author} {\bibfnamefont {V.~G.~J.}\ \bibnamefont
  {Stoks}}, \ and\ \bibinfo {author} {\bibfnamefont {R.}~\bibnamefont
  {Schiavilla}},\ }\href {\doibase 10.1103/PhysRevC.51.38} {\bibfield
  {journal} {\bibinfo  {journal} {Phys. Rev. C}\ }\textbf {\bibinfo {volume}
  {51}},\ \bibinfo {pages} {38} (\bibinfo {year} {1995})}\BibitemShut {NoStop}%
\bibitem [{\citenamefont {Machleidt}(2001)}]{machleidt01}%
  \BibitemOpen
  \bibfield  {author} {\bibinfo {author} {\bibfnamefont {R.}~\bibnamefont
  {Machleidt}},\ }\href {\doibase 10.1103/PhysRevC.63.024001} {\bibfield
  {journal} {\bibinfo  {journal} {Phys. Rev. C}\ }\textbf {\bibinfo {volume}
  {63}},\ \bibinfo {pages} {024001} (\bibinfo {year} {2001})}\BibitemShut
  {NoStop}%
\bibitem [{\citenamefont {Entem}\ and\ \citenamefont
  {Machleidt}(2003)}]{entem03}%
  \BibitemOpen
  \bibfield  {author} {\bibinfo {author} {\bibfnamefont {D.~R.}\ \bibnamefont
  {Entem}}\ and\ \bibinfo {author} {\bibfnamefont {R.}~\bibnamefont
  {Machleidt}},\ }\href {\doibase 10.1103/PhysRevC.68.041001} {\bibfield
  {journal} {\bibinfo  {journal} {Phys. Rev. C}\ }\textbf {\bibinfo {volume}
  {68}},\ \bibinfo {pages} {041001} (\bibinfo {year} {2003})}\BibitemShut
  {NoStop}%
\bibitem [{\citenamefont {Epelbaum}\ \emph {et~al.}(2005)\citenamefont
  {Epelbaum}, \citenamefont {Gl{\"o}ckle},\ and\ \citenamefont
  {Mei{\ss}ner}}]{epelbaum05}%
  \BibitemOpen
  \bibfield  {author} {\bibinfo {author} {\bibfnamefont {E.}~\bibnamefont
  {Epelbaum}}, \bibinfo {author} {\bibfnamefont {W.}~\bibnamefont
  {Gl{\"o}ckle}}, \ and\ \bibinfo {author} {\bibfnamefont {U.-G.}\ \bibnamefont
  {Mei{\ss}ner}},\ }\href {\doibase 10.1016/j.nuclphysa.2004.09.107} {\bibfield
   {journal} {\bibinfo  {journal} {Nucl. Phys. A}\ }\textbf {\bibinfo {volume}
  {747}},\ \bibinfo {pages} {362 } (\bibinfo {year} {2005})}\BibitemShut
  {NoStop}%
\bibitem [{\citenamefont {Benhar}\ \emph {et~al.}(1994)\citenamefont {Benhar},
  \citenamefont {Fabrocini}, \citenamefont {Fantoni},\ and\ \citenamefont
  {Sick}}]{benhar94}%
  \BibitemOpen
  \bibfield  {author} {\bibinfo {author} {\bibfnamefont {O.}~\bibnamefont
  {Benhar}}, \bibinfo {author} {\bibfnamefont {A.}~\bibnamefont {Fabrocini}},
  \bibinfo {author} {\bibfnamefont {S.}~\bibnamefont {Fantoni}}, \ and\
  \bibinfo {author} {\bibfnamefont {I.}~\bibnamefont {Sick}},\ }\href {\doibase
  10.1016/0375-9474(94)90920-2} {\bibfield  {journal} {\bibinfo  {journal}
  {Nucl. Phys. A}\ }\textbf {\bibinfo {volume} {579}},\ \bibinfo {pages} {493 }
  (\bibinfo {year} {1994})}\BibitemShut {NoStop}%
\bibitem [{\citenamefont {Pandharipande}\ \emph {et~al.}(1997)\citenamefont
  {Pandharipande}, \citenamefont {Sick},\ and\ \citenamefont
  {Huberts}}]{pandharipande97}%
  \BibitemOpen
  \bibfield  {author} {\bibinfo {author} {\bibfnamefont {V.~R.}\ \bibnamefont
  {Pandharipande}}, \bibinfo {author} {\bibfnamefont {I.}~\bibnamefont {Sick}},
  \ and\ \bibinfo {author} {\bibfnamefont {P.~K. A.~d.}\ \bibnamefont
  {Huberts}},\ }\href {\doibase 10.1103/RevModPhys.69.981} {\bibfield
  {journal} {\bibinfo  {journal} {Rev. Mod. Phys.}\ }\textbf {\bibinfo {volume}
  {69}},\ \bibinfo {pages} {981} (\bibinfo {year} {1997})}\BibitemShut
  {NoStop}%
\bibitem [{\citenamefont {M\"uther}\ and\ \citenamefont
  {Polls}(2000)}]{muether00}%
  \BibitemOpen
  \bibfield  {author} {\bibinfo {author} {\bibfnamefont {H.}~\bibnamefont
  {M\"uther}}\ and\ \bibinfo {author} {\bibfnamefont {A.}~\bibnamefont
  {Polls}},\ }\href {\doibase 10.1016/S0146-6410(00)00105-8} {\bibfield
  {journal} {\bibinfo  {journal} {Prog. Part. Nucl. Phys.}\ }\textbf {\bibinfo
  {volume} {45}},\ \bibinfo {pages} {243} (\bibinfo {year} {2000})}\BibitemShut
  {NoStop}%
\bibitem [{\citenamefont {Neff}\ and\ \citenamefont
  {Feldmeier}(2003)}]{ucom03}%
  \BibitemOpen
  \bibfield  {author} {\bibinfo {author} {\bibfnamefont {T.}~\bibnamefont
  {Neff}}\ and\ \bibinfo {author} {\bibfnamefont {H.}~\bibnamefont
  {Feldmeier}},\ }\href {\doibase 10.1016/S0375-9474(02)01307-6} {\bibfield
  {journal} {\bibinfo  {journal} {Nucl. Phys. A}\ }\textbf {\bibinfo {volume}
  {713}},\ \bibinfo {pages} {311 } (\bibinfo {year} {2003})}\BibitemShut
  {NoStop}%
\bibitem [{\citenamefont {Dickhoff}\ and\ \citenamefont
  {Barbieri}(2004)}]{dickhoff04}%
  \BibitemOpen
  \bibfield  {author} {\bibinfo {author} {\bibfnamefont {W.}~\bibnamefont
  {Dickhoff}}\ and\ \bibinfo {author} {\bibfnamefont {C.}~\bibnamefont
  {Barbieri}},\ }\href {\doibase 10.1016/j.ppnp.2004.02.038} {\bibfield
  {journal} {\bibinfo  {journal} {Prog. Part. Nucl. Phys.}\ }\textbf {\bibinfo
  {volume} {52}},\ \bibinfo {pages} {377 } (\bibinfo {year}
  {2004})}\BibitemShut {NoStop}%
\bibitem [{\citenamefont {Rios}\ \emph {et~al.}(2009)\citenamefont {Rios},
  \citenamefont {Polls},\ and\ \citenamefont {Dickhoff}}]{rios09}%
  \BibitemOpen
  \bibfield  {author} {\bibinfo {author} {\bibfnamefont {A.}~\bibnamefont
  {Rios}}, \bibinfo {author} {\bibfnamefont {A.}~\bibnamefont {Polls}}, \ and\
  \bibinfo {author} {\bibfnamefont {W.~H.}\ \bibnamefont {Dickhoff}},\ }\href
  {\doibase 10.1103/PhysRevC.79.064308} {\bibfield  {journal} {\bibinfo
  {journal} {Phys. Rev. C}\ }\textbf {\bibinfo {volume} {79}},\ \bibinfo
  {pages} {064308} (\bibinfo {year} {2009})}\BibitemShut {NoStop}%
\bibitem [{\citenamefont {Frankfurt}\ \emph {et~al.}(2008)\citenamefont
  {Frankfurt}, \citenamefont {Sargsian},\ and\ \citenamefont
  {Strikman}}]{frankfurt08}%
  \BibitemOpen
  \bibfield  {author} {\bibinfo {author} {\bibfnamefont {L.}~\bibnamefont
  {Frankfurt}}, \bibinfo {author} {\bibfnamefont {M.}~\bibnamefont {Sargsian}},
  \ and\ \bibinfo {author} {\bibfnamefont {M.}~\bibnamefont {Strikman}},\
  }\href {\doibase 10.1142/S0217751X08041207} {\bibfield  {journal} {\bibinfo
  {journal} {Int. J. Mod. Phys. A}\ }\textbf {\bibinfo {volume} {23}},\
  \bibinfo {pages} {2991} (\bibinfo {year} {2008})}\BibitemShut {NoStop}%
\bibitem [{\citenamefont {Arrington}\ \emph {et~al.}(2012)\citenamefont
  {Arrington}, \citenamefont {Higinbotham}, \citenamefont {Rosner},\ and\
  \citenamefont {Sargsian}}]{arrington12}%
  \BibitemOpen
  \bibfield  {author} {\bibinfo {author} {\bibfnamefont {J.}~\bibnamefont
  {Arrington}}, \bibinfo {author} {\bibfnamefont {D.}~\bibnamefont
  {Higinbotham}}, \bibinfo {author} {\bibfnamefont {G.}~\bibnamefont {Rosner}},
  \ and\ \bibinfo {author} {\bibfnamefont {M.}~\bibnamefont {Sargsian}},\
  }\href {\doibase 10.1016/j.ppnp.2012.04.002} {\bibfield  {journal} {\bibinfo
  {journal} {Prog. Part. Nucl. Phys.}\ }\textbf {\bibinfo {volume} {67}},\
  \bibinfo {pages} {898 } (\bibinfo {year} {2012})}\BibitemShut {NoStop}%
\bibitem [{\citenamefont {Tang}\ \emph {et~al.}(2003)\citenamefont {Tang},
  \citenamefont {Watson}, \citenamefont {Aclander}, \citenamefont {Alster},
  \citenamefont {Asryan}, \citenamefont {Averichev}, \citenamefont {Barton},
  \citenamefont {Baturin}, \citenamefont {Bukhtoyarova}, \citenamefont
  {Carroll}, \citenamefont {Gushue}, \citenamefont {Heppelmann}, \citenamefont
  {Leksanov}, \citenamefont {Makdisi}, \citenamefont {Malki}, \citenamefont
  {Minina}, \citenamefont {Navon}, \citenamefont {Nicholson}, \citenamefont
  {Ogawa}, \citenamefont {Panebratsev}, \citenamefont {Piasetzky},
  \citenamefont {Schetkovsky}, \citenamefont {Shimanskiy},\ and\ \citenamefont
  {Zhalov}}]{tang03}%
  \BibitemOpen
  \bibfield  {author} {\bibinfo {author} {\bibfnamefont {A.}~\bibnamefont
  {Tang}}, \bibinfo {author} {\bibfnamefont {J.~W.}\ \bibnamefont {Watson}},
  \bibinfo {author} {\bibfnamefont {J.}~\bibnamefont {Aclander}}, \bibinfo
  {author} {\bibfnamefont {J.}~\bibnamefont {Alster}}, \bibinfo {author}
  {\bibfnamefont {G.}~\bibnamefont {Asryan}}, \bibinfo {author} {\bibfnamefont
  {Y.}~\bibnamefont {Averichev}}, \bibinfo {author} {\bibfnamefont
  {D.}~\bibnamefont {Barton}}, \bibinfo {author} {\bibfnamefont
  {V.}~\bibnamefont {Baturin}}, \bibinfo {author} {\bibfnamefont
  {N.}~\bibnamefont {Bukhtoyarova}}, \bibinfo {author} {\bibfnamefont
  {A.}~\bibnamefont {Carroll}}, \bibinfo {author} {\bibfnamefont
  {S.}~\bibnamefont {Gushue}}, \bibinfo {author} {\bibfnamefont
  {S.}~\bibnamefont {Heppelmann}}, \bibinfo {author} {\bibfnamefont
  {A.}~\bibnamefont {Leksanov}}, \bibinfo {author} {\bibfnamefont
  {Y.}~\bibnamefont {Makdisi}}, \bibinfo {author} {\bibfnamefont
  {A.}~\bibnamefont {Malki}}, \bibinfo {author} {\bibfnamefont
  {E.}~\bibnamefont {Minina}}, \bibinfo {author} {\bibfnamefont
  {I.}~\bibnamefont {Navon}}, \bibinfo {author} {\bibfnamefont
  {H.}~\bibnamefont {Nicholson}}, \bibinfo {author} {\bibfnamefont
  {A.}~\bibnamefont {Ogawa}}, \bibinfo {author} {\bibfnamefont
  {Y.}~\bibnamefont {Panebratsev}}, \bibinfo {author} {\bibfnamefont
  {E.}~\bibnamefont {Piasetzky}}, \bibinfo {author} {\bibfnamefont
  {A.}~\bibnamefont {Schetkovsky}}, \bibinfo {author} {\bibfnamefont
  {S.}~\bibnamefont {Shimanskiy}}, \ and\ \bibinfo {author} {\bibfnamefont
  {D.}~\bibnamefont {Zhalov}},\ }\href {\doibase 10.1103/PhysRevLett.90.042301}
  {\bibfield  {journal} {\bibinfo  {journal} {Phys. Rev. Lett.}\ }\textbf
  {\bibinfo {volume} {90}},\ \bibinfo {pages} {042301} (\bibinfo {year}
  {2003})}\BibitemShut {NoStop}%
\bibitem [{\citenamefont {Shneor}\ \emph {et~al.}(2007)\citenamefont {Shneor}
  \emph {et~al.}}]{shneor07}%
  \BibitemOpen
  \bibfield  {author} {\bibinfo {author} {\bibfnamefont {R.}~\bibnamefont
  {Shneor}} \emph {et~al.} (\bibinfo {collaboration} {Jefferson Lab Hall A
  Collaboration}),\ }\href {\doibase 10.1103/PhysRevLett.99.072501} {\bibfield
  {journal} {\bibinfo  {journal} {Phys. Rev. Lett.}\ }\textbf {\bibinfo
  {volume} {99}},\ \bibinfo {pages} {072501} (\bibinfo {year}
  {2007})}\BibitemShut {NoStop}%
\bibitem [{\citenamefont {Subedi}\ \emph {et~al.}(2008)\citenamefont {Subedi},
  \citenamefont {Shneor}, \citenamefont {Monaghan}, \citenamefont {Anderson},
  \citenamefont {Aniol}, \citenamefont {Annand}, \citenamefont {Arrington},
  \citenamefont {Benaoum}, \citenamefont {Benmokhtar}, \citenamefont {Boeglin},
  \citenamefont {Chen}, \citenamefont {Choi}, \citenamefont {Cisbani},
  \citenamefont {Craver}, \citenamefont {Frullani}, \citenamefont {Garibaldi},
  \citenamefont {Gilad}, \citenamefont {Gilman}, \citenamefont {Glamazdin},
  \citenamefont {Hansen}, \citenamefont {Higinbotham}, \citenamefont
  {Holmstrom}, \citenamefont {Ibrahim}, \citenamefont {Igarashi}, \citenamefont
  {de~Jager}, \citenamefont {Jans}, \citenamefont {Jiang}, \citenamefont
  {Kaufman}, \citenamefont {Kelleher}, \citenamefont {Kolarkar}, \citenamefont
  {Kumbartzki}, \citenamefont {LeRose}, \citenamefont {Lindgren}, \citenamefont
  {Liyanage}, \citenamefont {Margaziotis}, \citenamefont {Markowitz},
  \citenamefont {Marrone}, \citenamefont {Mazouz}, \citenamefont {Meekins},
  \citenamefont {Michaels}, \citenamefont {Moffit}, \citenamefont {Perdrisat},
  \citenamefont {Piasetzky}, \citenamefont {Potokar}, \citenamefont {Punjabi},
  \citenamefont {Qiang}, \citenamefont {Reinhold}, \citenamefont {Ron},
  \citenamefont {Rosner}, \citenamefont {Saha}, \citenamefont {Sawatzky},
  \citenamefont {Shahinyan}, \citenamefont {{\v S}irca}, \citenamefont
  {Slifer}, \citenamefont {Solvignon}, \citenamefont {Sulkosky}, \citenamefont
  {Urciuoli}, \citenamefont {Voutier}, \citenamefont {Watson}, \citenamefont
  {Weinstein}, \citenamefont {Wojtsekhowski}, \citenamefont {Wood},
  \citenamefont {Zheng},\ and\ \citenamefont {Zhu}}]{subedi08}%
  \BibitemOpen
  \bibfield  {author} {\bibinfo {author} {\bibfnamefont {R.}~\bibnamefont
  {Subedi}}, \bibinfo {author} {\bibfnamefont {R.}~\bibnamefont {Shneor}},
  \bibinfo {author} {\bibfnamefont {P.}~\bibnamefont {Monaghan}}, \bibinfo
  {author} {\bibfnamefont {B.~D.}\ \bibnamefont {Anderson}}, \bibinfo {author}
  {\bibfnamefont {K.}~\bibnamefont {Aniol}}, \bibinfo {author} {\bibfnamefont
  {J.}~\bibnamefont {Annand}}, \bibinfo {author} {\bibfnamefont
  {J.}~\bibnamefont {Arrington}}, \bibinfo {author} {\bibfnamefont
  {H.}~\bibnamefont {Benaoum}}, \bibinfo {author} {\bibfnamefont
  {F.}~\bibnamefont {Benmokhtar}}, \bibinfo {author} {\bibfnamefont
  {W.}~\bibnamefont {Boeglin}}, \bibinfo {author} {\bibfnamefont {J.-P.}\
  \bibnamefont {Chen}}, \bibinfo {author} {\bibfnamefont {S.}~\bibnamefont
  {Choi}}, \bibinfo {author} {\bibfnamefont {E.}~\bibnamefont {Cisbani}},
  \bibinfo {author} {\bibfnamefont {B.}~\bibnamefont {Craver}}, \bibinfo
  {author} {\bibfnamefont {S.}~\bibnamefont {Frullani}}, \bibinfo {author}
  {\bibfnamefont {F.}~\bibnamefont {Garibaldi}}, \bibinfo {author}
  {\bibfnamefont {S.}~\bibnamefont {Gilad}}, \bibinfo {author} {\bibfnamefont
  {R.}~\bibnamefont {Gilman}}, \bibinfo {author} {\bibfnamefont
  {O.}~\bibnamefont {Glamazdin}}, \bibinfo {author} {\bibfnamefont {J.-O.}\
  \bibnamefont {Hansen}}, \bibinfo {author} {\bibfnamefont {D.~W.}\
  \bibnamefont {Higinbotham}}, \bibinfo {author} {\bibfnamefont
  {T.}~\bibnamefont {Holmstrom}}, \bibinfo {author} {\bibfnamefont
  {H.}~\bibnamefont {Ibrahim}}, \bibinfo {author} {\bibfnamefont
  {R.}~\bibnamefont {Igarashi}}, \bibinfo {author} {\bibfnamefont {C.~W.}\
  \bibnamefont {de~Jager}}, \bibinfo {author} {\bibfnamefont {E.}~\bibnamefont
  {Jans}}, \bibinfo {author} {\bibfnamefont {X.}~\bibnamefont {Jiang}},
  \bibinfo {author} {\bibfnamefont {L.~J.}\ \bibnamefont {Kaufman}}, \bibinfo
  {author} {\bibfnamefont {A.}~\bibnamefont {Kelleher}}, \bibinfo {author}
  {\bibfnamefont {A.}~\bibnamefont {Kolarkar}}, \bibinfo {author}
  {\bibfnamefont {G.}~\bibnamefont {Kumbartzki}}, \bibinfo {author}
  {\bibfnamefont {J.~J.}\ \bibnamefont {LeRose}}, \bibinfo {author}
  {\bibfnamefont {R.}~\bibnamefont {Lindgren}}, \bibinfo {author}
  {\bibfnamefont {N.}~\bibnamefont {Liyanage}}, \bibinfo {author}
  {\bibfnamefont {D.~J.}\ \bibnamefont {Margaziotis}}, \bibinfo {author}
  {\bibfnamefont {P.}~\bibnamefont {Markowitz}}, \bibinfo {author}
  {\bibfnamefont {S.}~\bibnamefont {Marrone}}, \bibinfo {author} {\bibfnamefont
  {M.}~\bibnamefont {Mazouz}}, \bibinfo {author} {\bibfnamefont
  {D.}~\bibnamefont {Meekins}}, \bibinfo {author} {\bibfnamefont
  {R.}~\bibnamefont {Michaels}}, \bibinfo {author} {\bibfnamefont
  {B.}~\bibnamefont {Moffit}}, \bibinfo {author} {\bibfnamefont {C.~F.}\
  \bibnamefont {Perdrisat}}, \bibinfo {author} {\bibfnamefont {E.}~\bibnamefont
  {Piasetzky}}, \bibinfo {author} {\bibfnamefont {M.}~\bibnamefont {Potokar}},
  \bibinfo {author} {\bibfnamefont {V.}~\bibnamefont {Punjabi}}, \bibinfo
  {author} {\bibfnamefont {Y.}~\bibnamefont {Qiang}}, \bibinfo {author}
  {\bibfnamefont {J.}~\bibnamefont {Reinhold}}, \bibinfo {author}
  {\bibfnamefont {G.}~\bibnamefont {Ron}}, \bibinfo {author} {\bibfnamefont
  {G.}~\bibnamefont {Rosner}}, \bibinfo {author} {\bibfnamefont
  {A.}~\bibnamefont {Saha}}, \bibinfo {author} {\bibfnamefont {B.}~\bibnamefont
  {Sawatzky}}, \bibinfo {author} {\bibfnamefont {A.}~\bibnamefont {Shahinyan}},
  \bibinfo {author} {\bibfnamefont {S.}~\bibnamefont {{\v S}irca}}, \bibinfo
  {author} {\bibfnamefont {K.}~\bibnamefont {Slifer}}, \bibinfo {author}
  {\bibfnamefont {P.}~\bibnamefont {Solvignon}}, \bibinfo {author}
  {\bibfnamefont {V.}~\bibnamefont {Sulkosky}}, \bibinfo {author}
  {\bibfnamefont {G.~M.}\ \bibnamefont {Urciuoli}}, \bibinfo {author}
  {\bibfnamefont {E.}~\bibnamefont {Voutier}}, \bibinfo {author} {\bibfnamefont
  {J.~W.}\ \bibnamefont {Watson}}, \bibinfo {author} {\bibfnamefont {L.~B.}\
  \bibnamefont {Weinstein}}, \bibinfo {author} {\bibfnamefont {B.}~\bibnamefont
  {Wojtsekhowski}}, \bibinfo {author} {\bibfnamefont {S.}~\bibnamefont {Wood}},
  \bibinfo {author} {\bibfnamefont {X.-C.}\ \bibnamefont {Zheng}}, \ and\
  \bibinfo {author} {\bibfnamefont {L.}~\bibnamefont {Zhu}},\ }\href {\doibase
  10.1126/science.1156675} {\bibfield  {journal} {\bibinfo  {journal}
  {Science}\ }\textbf {\bibinfo {volume} {320}},\ \bibinfo {pages} {1476}
  (\bibinfo {year} {2008})}\BibitemShut {NoStop}%
\bibitem [{\citenamefont {Baghdasaryan}\ \emph {et~al.}(2010)\citenamefont
  {Baghdasaryan} \emph {et~al.}}]{badhdasaryan10}%
  \BibitemOpen
  \bibfield  {author} {\bibinfo {author} {\bibfnamefont {H.}~\bibnamefont
  {Baghdasaryan}} \emph {et~al.} (\bibinfo {collaboration} {CLAS
  Collaboration}),\ }\href {\doibase 10.1103/PhysRevLett.105.222501} {\bibfield
   {journal} {\bibinfo  {journal} {Phys. Rev. Lett.}\ }\textbf {\bibinfo
  {volume} {105}},\ \bibinfo {pages} {222501} (\bibinfo {year}
  {2010})}\BibitemShut {NoStop}%
\bibitem [{\citenamefont {Korover}\ \emph {et~al.}(2014)\citenamefont {Korover}
  \emph {et~al.}}]{korover14}%
  \BibitemOpen
  \bibfield  {author} {\bibinfo {author} {\bibfnamefont {I.}~\bibnamefont
  {Korover}} \emph {et~al.} (\bibinfo {collaboration} {Jefferson Lab Hall A
  Collaboration}),\ }\href {\doibase 10.1103/PhysRevLett.113.022501} {\bibfield
   {journal} {\bibinfo  {journal} {Phys. Rev. Lett.}\ }\textbf {\bibinfo
  {volume} {113}},\ \bibinfo {pages} {022501} (\bibinfo {year}
  {2014})}\BibitemShut {NoStop}%
\bibitem [{\citenamefont {Miki}\ \emph {et~al.}(2013)\citenamefont {Miki},
  \citenamefont {Tamii}, \citenamefont {Aoi}, \citenamefont {Fukui},
  \citenamefont {Hashimoto}, \citenamefont {Hatanaka}, \citenamefont {Ito},
  \citenamefont {Kawabata}, \citenamefont {Matsubara}, \citenamefont {Ogata},
  \citenamefont {Ong}, \citenamefont {Sakaguchi}, \citenamefont {Sakaguchi},
  \citenamefont {Suzuki}, \citenamefont {Tanaka}, \citenamefont {Tanihata},
  \citenamefont {Uesaka},\ and\ \citenamefont {Yamamoto}}]{miki13}%
  \BibitemOpen
  \bibfield  {author} {\bibinfo {author} {\bibfnamefont {K.}~\bibnamefont
  {Miki}}, \bibinfo {author} {\bibfnamefont {A.}~\bibnamefont {Tamii}},
  \bibinfo {author} {\bibfnamefont {N.}~\bibnamefont {Aoi}}, \bibinfo {author}
  {\bibfnamefont {T.}~\bibnamefont {Fukui}}, \bibinfo {author} {\bibfnamefont
  {T.}~\bibnamefont {Hashimoto}}, \bibinfo {author} {\bibfnamefont
  {K.}~\bibnamefont {Hatanaka}}, \bibinfo {author} {\bibfnamefont
  {T.}~\bibnamefont {Ito}}, \bibinfo {author} {\bibfnamefont {T.}~\bibnamefont
  {Kawabata}}, \bibinfo {author} {\bibfnamefont {H.}~\bibnamefont {Matsubara}},
  \bibinfo {author} {\bibfnamefont {K.}~\bibnamefont {Ogata}}, \bibinfo
  {author} {\bibfnamefont {H.}~\bibnamefont {Ong}}, \bibinfo {author}
  {\bibfnamefont {H.}~\bibnamefont {Sakaguchi}}, \bibinfo {author}
  {\bibfnamefont {S.}~\bibnamefont {Sakaguchi}}, \bibinfo {author}
  {\bibfnamefont {T.}~\bibnamefont {Suzuki}}, \bibinfo {author} {\bibfnamefont
  {J.}~\bibnamefont {Tanaka}}, \bibinfo {author} {\bibfnamefont
  {I.}~\bibnamefont {Tanihata}}, \bibinfo {author} {\bibfnamefont
  {T.}~\bibnamefont {Uesaka}}, \ and\ \bibinfo {author} {\bibfnamefont
  {T.}~\bibnamefont {Yamamoto}},\ }\href {\doibase 10.1007/s00601-013-0611-7}
  {\bibfield  {journal} {\bibinfo  {journal} {Few-Body Syst.}\ }\textbf
  {\bibinfo {volume} {54}},\ \bibinfo {pages} {1353} (\bibinfo {year}
  {2013})}\BibitemShut {NoStop}%
\bibitem [{\citenamefont {Ong}\ \emph {et~al.}(2013)\citenamefont {Ong},
  \citenamefont {Tanihata}, \citenamefont {Tamii}, \citenamefont {Myo},
  \citenamefont {Ogata}, \citenamefont {Fukuda}, \citenamefont {Hirota},
  \citenamefont {Ikeda}, \citenamefont {Ishikawa}, \citenamefont {Kawabata},
  \citenamefont {Matsubara}, \citenamefont {Matsuta}, \citenamefont {Mihara},
  \citenamefont {Naito}, \citenamefont {Nishimura}, \citenamefont {Ogawa},
  \citenamefont {Okamura}, \citenamefont {Ozawa}, \citenamefont {Pang},
  \citenamefont {Sakaguchi}, \citenamefont {Sekiguchi}, \citenamefont {Suzuki},
  \citenamefont {Taniguchi}, \citenamefont {Takashina}, \citenamefont {Toki},
  \citenamefont {Yasuda}, \citenamefont {Yosoi},\ and\ \citenamefont
  {Zenihiro}}]{ong13}%
  \BibitemOpen
  \bibfield  {author} {\bibinfo {author} {\bibfnamefont {H.}~\bibnamefont
  {Ong}}, \bibinfo {author} {\bibfnamefont {I.}~\bibnamefont {Tanihata}},
  \bibinfo {author} {\bibfnamefont {A.}~\bibnamefont {Tamii}}, \bibinfo
  {author} {\bibfnamefont {T.}~\bibnamefont {Myo}}, \bibinfo {author}
  {\bibfnamefont {K.}~\bibnamefont {Ogata}}, \bibinfo {author} {\bibfnamefont
  {M.}~\bibnamefont {Fukuda}}, \bibinfo {author} {\bibfnamefont
  {K.}~\bibnamefont {Hirota}}, \bibinfo {author} {\bibfnamefont
  {K.}~\bibnamefont {Ikeda}}, \bibinfo {author} {\bibfnamefont
  {D.}~\bibnamefont {Ishikawa}}, \bibinfo {author} {\bibfnamefont
  {T.}~\bibnamefont {Kawabata}}, \bibinfo {author} {\bibfnamefont
  {H.}~\bibnamefont {Matsubara}}, \bibinfo {author} {\bibfnamefont
  {K.}~\bibnamefont {Matsuta}}, \bibinfo {author} {\bibfnamefont
  {M.}~\bibnamefont {Mihara}}, \bibinfo {author} {\bibfnamefont
  {T.}~\bibnamefont {Naito}}, \bibinfo {author} {\bibfnamefont
  {D.}~\bibnamefont {Nishimura}}, \bibinfo {author} {\bibfnamefont
  {Y.}~\bibnamefont {Ogawa}}, \bibinfo {author} {\bibfnamefont
  {H.}~\bibnamefont {Okamura}}, \bibinfo {author} {\bibfnamefont
  {A.}~\bibnamefont {Ozawa}}, \bibinfo {author} {\bibfnamefont
  {D.}~\bibnamefont {Pang}}, \bibinfo {author} {\bibfnamefont {H.}~\bibnamefont
  {Sakaguchi}}, \bibinfo {author} {\bibfnamefont {K.}~\bibnamefont
  {Sekiguchi}}, \bibinfo {author} {\bibfnamefont {T.}~\bibnamefont {Suzuki}},
  \bibinfo {author} {\bibfnamefont {M.}~\bibnamefont {Taniguchi}}, \bibinfo
  {author} {\bibfnamefont {M.}~\bibnamefont {Takashina}}, \bibinfo {author}
  {\bibfnamefont {H.}~\bibnamefont {Toki}}, \bibinfo {author} {\bibfnamefont
  {Y.}~\bibnamefont {Yasuda}}, \bibinfo {author} {\bibfnamefont
  {M.}~\bibnamefont {Yosoi}}, \ and\ \bibinfo {author} {\bibfnamefont
  {J.}~\bibnamefont {Zenihiro}},\ }\href {\doibase
  10.1016/j.physletb.2013.07.038} {\bibfield  {journal} {\bibinfo  {journal}
  {Phys. Lett. B}\ }\textbf {\bibinfo {volume} {725}},\ \bibinfo {pages} {277}
  (\bibinfo {year} {2013})}\BibitemShut {NoStop}%
\bibitem [{\citenamefont {Vanhalst}\ \emph {et~al.}(2012)\citenamefont
  {Vanhalst}, \citenamefont {Ryckebusch},\ and\ \citenamefont
  {Cosyn}}]{vanhalst12}%
  \BibitemOpen
  \bibfield  {author} {\bibinfo {author} {\bibfnamefont {M.}~\bibnamefont
  {Vanhalst}}, \bibinfo {author} {\bibfnamefont {J.}~\bibnamefont
  {Ryckebusch}}, \ and\ \bibinfo {author} {\bibfnamefont {W.}~\bibnamefont
  {Cosyn}},\ }\href {\doibase 10.1103/PhysRevC.86.044619} {\bibfield  {journal}
  {\bibinfo  {journal} {Phys. Rev. C}\ }\textbf {\bibinfo {volume} {86}},\
  \bibinfo {pages} {044619} (\bibinfo {year} {2012})}\BibitemShut {NoStop}%
\bibitem [{\citenamefont {Weiss}\ \emph {et~al.}()\citenamefont {Weiss},
  \citenamefont {Bazak},\ and\ \citenamefont {Barnea}}]{weiss15}%
  \BibitemOpen
  \bibfield  {author} {\bibinfo {author} {\bibfnamefont {R.}~\bibnamefont
  {Weiss}}, \bibinfo {author} {\bibfnamefont {B.}~\bibnamefont {Bazak}}, \ and\
  \bibinfo {author} {\bibfnamefont {N.}~\bibnamefont {Barnea}},\ }\href@noop {}
  {\ }\Eprint {http://arxiv.org/abs/1503.07047} {arXiv:1503.07047 [nucl-th]}
  \BibitemShut {NoStop}%
\bibitem [{\citenamefont {Schiavilla}\ \emph {et~al.}(2007)\citenamefont
  {Schiavilla}, \citenamefont {Wiringa}, \citenamefont {Pieper},\ and\
  \citenamefont {Carlson}}]{schiavilla07}%
  \BibitemOpen
  \bibfield  {author} {\bibinfo {author} {\bibfnamefont {R.}~\bibnamefont
  {Schiavilla}}, \bibinfo {author} {\bibfnamefont {R.~B.}\ \bibnamefont
  {Wiringa}}, \bibinfo {author} {\bibfnamefont {S.~C.}\ \bibnamefont {Pieper}},
  \ and\ \bibinfo {author} {\bibfnamefont {J.}~\bibnamefont {Carlson}},\ }\href
  {\doibase 10.1103/PhysRevLett.98.132501} {\bibfield  {journal} {\bibinfo
  {journal} {Phys. Rev. Lett.}\ }\textbf {\bibinfo {volume} {98}},\ \bibinfo
  {pages} {132501} (\bibinfo {year} {2007})}\BibitemShut {NoStop}%
\bibitem [{\citenamefont {Wiringa}\ \emph {et~al.}(2008)\citenamefont
  {Wiringa}, \citenamefont {Schiavilla}, \citenamefont {Pieper},\ and\
  \citenamefont {Carlson}}]{wiringa08}%
  \BibitemOpen
  \bibfield  {author} {\bibinfo {author} {\bibfnamefont {R.~B.}\ \bibnamefont
  {Wiringa}}, \bibinfo {author} {\bibfnamefont {R.}~\bibnamefont {Schiavilla}},
  \bibinfo {author} {\bibfnamefont {S.~C.}\ \bibnamefont {Pieper}}, \ and\
  \bibinfo {author} {\bibfnamefont {J.}~\bibnamefont {Carlson}},\ }\href
  {\doibase 10.1103/PhysRevC.78.021001} {\bibfield  {journal} {\bibinfo
  {journal} {Phys. Rev. C}\ }\textbf {\bibinfo {volume} {78}},\ \bibinfo
  {pages} {021001} (\bibinfo {year} {2008})}\BibitemShut {NoStop}%
\bibitem [{\citenamefont {Wiringa}\ \emph {et~al.}(2014)\citenamefont
  {Wiringa}, \citenamefont {Schiavilla}, \citenamefont {Pieper},\ and\
  \citenamefont {Carlson}}]{wiringa14}%
  \BibitemOpen
  \bibfield  {author} {\bibinfo {author} {\bibfnamefont {R.~B.}\ \bibnamefont
  {Wiringa}}, \bibinfo {author} {\bibfnamefont {R.}~\bibnamefont {Schiavilla}},
  \bibinfo {author} {\bibfnamefont {S.~C.}\ \bibnamefont {Pieper}}, \ and\
  \bibinfo {author} {\bibfnamefont {J.}~\bibnamefont {Carlson}},\ }\href
  {\doibase 10.1103/PhysRevC.89.024305} {\bibfield  {journal} {\bibinfo
  {journal} {Phys. Rev. C}\ }\textbf {\bibinfo {volume} {89}},\ \bibinfo
  {pages} {024305} (\bibinfo {year} {2014})}\BibitemShut {NoStop}%
\bibitem [{\citenamefont {Alvioli}\ \emph {et~al.}(2012)\citenamefont
  {Alvioli}, \citenamefont {Ciofi~degli Atti}, \citenamefont {Kaptari},
  \citenamefont {Mezzetti}, \citenamefont {Morita},\ and\ \citenamefont
  {Scopetta}}]{alvioli12}%
  \BibitemOpen
  \bibfield  {author} {\bibinfo {author} {\bibfnamefont {M.}~\bibnamefont
  {Alvioli}}, \bibinfo {author} {\bibfnamefont {C.}~\bibnamefont {Ciofi~degli
  Atti}}, \bibinfo {author} {\bibfnamefont {L.~P.}\ \bibnamefont {Kaptari}},
  \bibinfo {author} {\bibfnamefont {C.~B.}\ \bibnamefont {Mezzetti}}, \bibinfo
  {author} {\bibfnamefont {H.}~\bibnamefont {Morita}}, \ and\ \bibinfo {author}
  {\bibfnamefont {S.}~\bibnamefont {Scopetta}},\ }\href {\doibase
  10.1103/PhysRevC.85.021001} {\bibfield  {journal} {\bibinfo  {journal} {Phys.
  Rev. C}\ }\textbf {\bibinfo {volume} {85}},\ \bibinfo {pages} {021001}
  (\bibinfo {year} {2012})}\BibitemShut {NoStop}%
\bibitem [{\citenamefont {Alvioli}\ \emph
  {et~al.}(2013{\natexlab{a}})\citenamefont {Alvioli}, \citenamefont
  {Ciofi~degli Atti}, \citenamefont {Kaptari}, \citenamefont {Mezzetti},\ and\
  \citenamefont {Morita}}]{alvioli13a}%
  \BibitemOpen
  \bibfield  {author} {\bibinfo {author} {\bibfnamefont {M.}~\bibnamefont
  {Alvioli}}, \bibinfo {author} {\bibfnamefont {C.}~\bibnamefont {Ciofi~degli
  Atti}}, \bibinfo {author} {\bibfnamefont {L.~P.}\ \bibnamefont {Kaptari}},
  \bibinfo {author} {\bibfnamefont {C.~B.}\ \bibnamefont {Mezzetti}}, \ and\
  \bibinfo {author} {\bibfnamefont {H.}~\bibnamefont {Morita}},\ }\href
  {\doibase 10.1142/S021830131330021X} {\bibfield  {journal} {\bibinfo
  {journal} {Int. J. Mod. Phys. E}\ }\textbf {\bibinfo {volume} {22}},\
  \bibinfo {pages} {1330021} (\bibinfo {year}
  {2013}{\natexlab{a}})}\BibitemShut {NoStop}%
\bibitem [{\citenamefont {Fabrocini}\ \emph {et~al.}(2000)\citenamefont
  {Fabrocini}, \citenamefont {Arias~de Saavedra},\ and\ \citenamefont
  {Co'}}]{fabrocini00}%
  \BibitemOpen
  \bibfield  {author} {\bibinfo {author} {\bibfnamefont {A.}~\bibnamefont
  {Fabrocini}}, \bibinfo {author} {\bibfnamefont {F.}~\bibnamefont {Arias~de
  Saavedra}}, \ and\ \bibinfo {author} {\bibfnamefont {G.}~\bibnamefont
  {Co'}},\ }\href {\doibase 10.1103/PhysRevC.61.044302} {\bibfield  {journal}
  {\bibinfo  {journal} {Phys. Rev. C}\ }\textbf {\bibinfo {volume} {61}},\
  \bibinfo {pages} {044302} (\bibinfo {year} {2000})}\BibitemShut {NoStop}%
\bibitem [{\citenamefont {Bogner}\ \emph {et~al.}(2003)\citenamefont {Bogner},
  \citenamefont {Kuo},\ and\ \citenamefont {Schwenk}}]{bogner03}%
  \BibitemOpen
  \bibfield  {author} {\bibinfo {author} {\bibfnamefont {S.}~\bibnamefont
  {Bogner}}, \bibinfo {author} {\bibfnamefont {T.}~\bibnamefont {Kuo}}, \ and\
  \bibinfo {author} {\bibfnamefont {A.}~\bibnamefont {Schwenk}},\ }\href
  {\doibase 10.1016/j.physrep.2003.07.001} {\bibfield  {journal} {\bibinfo
  {journal} {Phys. Rep.}\ }\textbf {\bibinfo {volume} {386}},\ \bibinfo {pages}
  {1 } (\bibinfo {year} {2003})}\BibitemShut {NoStop}%
\bibitem [{\citenamefont {Feldmeier}\ \emph {et~al.}(1998)\citenamefont
  {Feldmeier}, \citenamefont {Neff}, \citenamefont {Roth},\ and\ \citenamefont
  {Schnack}}]{ucom98}%
  \BibitemOpen
  \bibfield  {author} {\bibinfo {author} {\bibfnamefont {H.}~\bibnamefont
  {Feldmeier}}, \bibinfo {author} {\bibfnamefont {T.}~\bibnamefont {Neff}},
  \bibinfo {author} {\bibfnamefont {R.}~\bibnamefont {Roth}}, \ and\ \bibinfo
  {author} {\bibfnamefont {J.}~\bibnamefont {Schnack}},\ }\href {\doibase
  10.1016/S0375-9474(97)00805-1} {\bibfield  {journal} {\bibinfo  {journal}
  {Nucl. Phys. A}\ }\textbf {\bibinfo {volume} {632}},\ \bibinfo {pages} {61 }
  (\bibinfo {year} {1998})}\BibitemShut {NoStop}%
\bibitem [{\citenamefont {Roth}\ \emph {et~al.}(2010)\citenamefont {Roth},
  \citenamefont {Neff},\ and\ \citenamefont {Feldmeier}}]{ucom10}%
  \BibitemOpen
  \bibfield  {author} {\bibinfo {author} {\bibfnamefont {R.}~\bibnamefont
  {Roth}}, \bibinfo {author} {\bibfnamefont {T.}~\bibnamefont {Neff}}, \ and\
  \bibinfo {author} {\bibfnamefont {H.}~\bibnamefont {Feldmeier}},\ }\href
  {\doibase 10.1016/j.ppnp.2010.02.003} {\bibfield  {journal} {\bibinfo
  {journal} {Prog. Part. Nucl. Phys.}\ }\textbf {\bibinfo {volume} {65}},\
  \bibinfo {pages} {50 } (\bibinfo {year} {2010})}\BibitemShut {NoStop}%
\bibitem [{\citenamefont {Bogner}\ \emph {et~al.}(2007)\citenamefont {Bogner},
  \citenamefont {Furnstahl},\ and\ \citenamefont {Perry}}]{bogner07}%
  \BibitemOpen
  \bibfield  {author} {\bibinfo {author} {\bibfnamefont {S.~K.}\ \bibnamefont
  {Bogner}}, \bibinfo {author} {\bibfnamefont {R.~J.}\ \bibnamefont
  {Furnstahl}}, \ and\ \bibinfo {author} {\bibfnamefont {R.~J.}\ \bibnamefont
  {Perry}},\ }\href {\doibase 10.1103/PhysRevC.75.061001} {\bibfield  {journal}
  {\bibinfo  {journal} {Phys. Rev. C}\ }\textbf {\bibinfo {volume} {75}},\
  \bibinfo {pages} {061001} (\bibinfo {year} {2007})}\BibitemShut {NoStop}%
\bibitem [{\citenamefont {Bogner}\ \emph {et~al.}(2010)\citenamefont {Bogner},
  \citenamefont {Furnstahl},\ and\ \citenamefont {Schwenk}}]{bogner10}%
  \BibitemOpen
  \bibfield  {author} {\bibinfo {author} {\bibfnamefont {S.}~\bibnamefont
  {Bogner}}, \bibinfo {author} {\bibfnamefont {R.}~\bibnamefont {Furnstahl}}, \
  and\ \bibinfo {author} {\bibfnamefont {A.}~\bibnamefont {Schwenk}},\ }\href
  {\doibase 10.1016/j.ppnp.2010.03.001} {\bibfield  {journal} {\bibinfo
  {journal} {Prog. Part. Nucl. Phys.}\ }\textbf {\bibinfo {volume} {65}},\
  \bibinfo {pages} {94 } (\bibinfo {year} {2010})}\BibitemShut {NoStop}%
\bibitem [{\citenamefont {Jurgenson}\ \emph {et~al.}(2009)\citenamefont
  {Jurgenson}, \citenamefont {Navr\'atil},\ and\ \citenamefont
  {Furnstahl}}]{jurgenson09}%
  \BibitemOpen
  \bibfield  {author} {\bibinfo {author} {\bibfnamefont {E.~D.}\ \bibnamefont
  {Jurgenson}}, \bibinfo {author} {\bibfnamefont {P.}~\bibnamefont
  {Navr\'atil}}, \ and\ \bibinfo {author} {\bibfnamefont {R.~J.}\ \bibnamefont
  {Furnstahl}},\ }\href {\doibase 10.1103/PhysRevLett.103.082501} {\bibfield
  {journal} {\bibinfo  {journal} {Phys. Rev. Lett.}\ }\textbf {\bibinfo
  {volume} {103}},\ \bibinfo {pages} {082501} (\bibinfo {year}
  {2009})}\BibitemShut {NoStop}%
\bibitem [{\citenamefont {Hebeler}(2012)}]{hebeler12}%
  \BibitemOpen
  \bibfield  {author} {\bibinfo {author} {\bibfnamefont {K.}~\bibnamefont
  {Hebeler}},\ }\href {\doibase 10.1103/PhysRevC.85.021002} {\bibfield
  {journal} {\bibinfo  {journal} {Phys. Rev. C}\ }\textbf {\bibinfo {volume}
  {85}},\ \bibinfo {pages} {021002} (\bibinfo {year} {2012})}\BibitemShut
  {NoStop}%
\bibitem [{\citenamefont {Roth}\ \emph {et~al.}(2014)\citenamefont {Roth},
  \citenamefont {Calci}, \citenamefont {Langhammer},\ and\ \citenamefont
  {Binder}}]{roth14}%
  \BibitemOpen
  \bibfield  {author} {\bibinfo {author} {\bibfnamefont {R.}~\bibnamefont
  {Roth}}, \bibinfo {author} {\bibfnamefont {A.}~\bibnamefont {Calci}},
  \bibinfo {author} {\bibfnamefont {J.}~\bibnamefont {Langhammer}}, \ and\
  \bibinfo {author} {\bibfnamefont {S.}~\bibnamefont {Binder}},\ }\href
  {\doibase 10.1103/PhysRevC.90.024325} {\bibfield  {journal} {\bibinfo
  {journal} {Phys. Rev. C}\ }\textbf {\bibinfo {volume} {90}},\ \bibinfo
  {pages} {024325} (\bibinfo {year} {2014})}\BibitemShut {NoStop}%
\bibitem [{\citenamefont {Schuster}\ \emph {et~al.}(2014)\citenamefont
  {Schuster}, \citenamefont {Quaglioni}, \citenamefont {Johnson}, \citenamefont
  {Jurgenson},\ and\ \citenamefont {Navr\'atil}}]{schuster14}%
  \BibitemOpen
  \bibfield  {author} {\bibinfo {author} {\bibfnamefont {M.~D.}\ \bibnamefont
  {Schuster}}, \bibinfo {author} {\bibfnamefont {S.}~\bibnamefont {Quaglioni}},
  \bibinfo {author} {\bibfnamefont {C.~W.}\ \bibnamefont {Johnson}}, \bibinfo
  {author} {\bibfnamefont {E.~D.}\ \bibnamefont {Jurgenson}}, \ and\ \bibinfo
  {author} {\bibfnamefont {P.}~\bibnamefont {Navr\'atil}},\ }\href {\doibase
  10.1103/PhysRevC.90.011301} {\bibfield  {journal} {\bibinfo  {journal} {Phys.
  Rev. C}\ }\textbf {\bibinfo {volume} {90}},\ \bibinfo {pages} {011301}
  (\bibinfo {year} {2014})}\BibitemShut {NoStop}%
\bibitem [{\citenamefont {Feldmeier}\ \emph {et~al.}(2011)\citenamefont
  {Feldmeier}, \citenamefont {Horiuchi}, \citenamefont {Neff},\ and\
  \citenamefont {Suzuki}}]{src11}%
  \BibitemOpen
  \bibfield  {author} {\bibinfo {author} {\bibfnamefont {H.}~\bibnamefont
  {Feldmeier}}, \bibinfo {author} {\bibfnamefont {W.}~\bibnamefont {Horiuchi}},
  \bibinfo {author} {\bibfnamefont {T.}~\bibnamefont {Neff}}, \ and\ \bibinfo
  {author} {\bibfnamefont {Y.}~\bibnamefont {Suzuki}},\ }\href {\doibase
  10.1103/PhysRevC.84.054003} {\bibfield  {journal} {\bibinfo  {journal} {Phys.
  Rev. C}\ }\textbf {\bibinfo {volume} {84}},\ \bibinfo {pages} {054003}
  (\bibinfo {year} {2011})}\BibitemShut {NoStop}%
\bibitem [{\citenamefont {Anderson}\ \emph {et~al.}(2010)\citenamefont
  {Anderson}, \citenamefont {Bogner}, \citenamefont {Furnstahl},\ and\
  \citenamefont {Perry}}]{anderson10}%
  \BibitemOpen
  \bibfield  {author} {\bibinfo {author} {\bibfnamefont {E.~R.}\ \bibnamefont
  {Anderson}}, \bibinfo {author} {\bibfnamefont {S.~K.}\ \bibnamefont
  {Bogner}}, \bibinfo {author} {\bibfnamefont {R.~J.}\ \bibnamefont
  {Furnstahl}}, \ and\ \bibinfo {author} {\bibfnamefont {R.~J.}\ \bibnamefont
  {Perry}},\ }\href {\doibase 10.1103/PhysRevC.82.054001} {\bibfield  {journal}
  {\bibinfo  {journal} {Phys. Rev. C}\ }\textbf {\bibinfo {volume} {82}},\
  \bibinfo {pages} {054001} (\bibinfo {year} {2010})}\BibitemShut {NoStop}%
\bibitem [{\citenamefont {Bogner}\ and\ \citenamefont
  {Roscher}(2012)}]{bogner12}%
  \BibitemOpen
  \bibfield  {author} {\bibinfo {author} {\bibfnamefont {S.~K.}\ \bibnamefont
  {Bogner}}\ and\ \bibinfo {author} {\bibfnamefont {D.}~\bibnamefont
  {Roscher}},\ }\href {\doibase 10.1103/PhysRevC.86.064304} {\bibfield
  {journal} {\bibinfo  {journal} {Phys. Rev. C}\ }\textbf {\bibinfo {volume}
  {86}},\ \bibinfo {pages} {064304} (\bibinfo {year} {2012})}\BibitemShut
  {NoStop}%
\bibitem [{\citenamefont {Suzuki}\ and\ \citenamefont
  {Horiuchi}(2009)}]{suzuki09}%
  \BibitemOpen
  \bibfield  {author} {\bibinfo {author} {\bibfnamefont {Y.}~\bibnamefont
  {Suzuki}}\ and\ \bibinfo {author} {\bibfnamefont {W.}~\bibnamefont
  {Horiuchi}},\ }\href {\doibase 10.1016/j.nuclphysa.2008.12.009} {\bibfield
  {journal} {\bibinfo  {journal} {Nucl. Phys. A}\ }\textbf {\bibinfo {volume}
  {818}},\ \bibinfo {pages} {188 } (\bibinfo {year} {2009})}\BibitemShut
  {NoStop}%
\bibitem [{\citenamefont {Mitroy}\ \emph {et~al.}(2013)\citenamefont {Mitroy},
  \citenamefont {Bubin}, \citenamefont {Horiuchi}, \citenamefont {Suzuki},
  \citenamefont {Adamowicz}, \citenamefont {Cencek}, \citenamefont {Szalewicz},
  \citenamefont {Komasa}, \citenamefont {Blume},\ and\ \citenamefont
  {Varga}}]{mitroy13}%
  \BibitemOpen
  \bibfield  {author} {\bibinfo {author} {\bibfnamefont {J.}~\bibnamefont
  {Mitroy}}, \bibinfo {author} {\bibfnamefont {S.}~\bibnamefont {Bubin}},
  \bibinfo {author} {\bibfnamefont {W.}~\bibnamefont {Horiuchi}}, \bibinfo
  {author} {\bibfnamefont {Y.}~\bibnamefont {Suzuki}}, \bibinfo {author}
  {\bibfnamefont {L.}~\bibnamefont {Adamowicz}}, \bibinfo {author}
  {\bibfnamefont {W.}~\bibnamefont {Cencek}}, \bibinfo {author} {\bibfnamefont
  {K.}~\bibnamefont {Szalewicz}}, \bibinfo {author} {\bibfnamefont
  {J.}~\bibnamefont {Komasa}}, \bibinfo {author} {\bibfnamefont
  {D.}~\bibnamefont {Blume}}, \ and\ \bibinfo {author} {\bibfnamefont
  {K.}~\bibnamefont {Varga}},\ }\href {\doibase 10.1103/RevModPhys.85.693}
  {\bibfield  {journal} {\bibinfo  {journal} {Rev. Mod. Phys.}\ }\textbf
  {\bibinfo {volume} {85}},\ \bibinfo {pages} {693} (\bibinfo {year}
  {2013})}\BibitemShut {NoStop}%
\bibitem [{\citenamefont {Caurier}\ and\ \citenamefont
  {Nowacki}(1999)}]{caurier99}%
  \BibitemOpen
  \bibfield  {author} {\bibinfo {author} {\bibfnamefont {E.}~\bibnamefont
  {Caurier}}\ and\ \bibinfo {author} {\bibfnamefont {F.}~\bibnamefont
  {Nowacki}},\ }\href {http://www.actaphys.uj.edu.pl/vol30/abs/v30p0705.htm}
  {\bibfield  {journal} {\bibinfo  {journal} {Acta Phys. Pol. B}\ }\textbf
  {\bibinfo {volume} {30}},\ \bibinfo {pages} {705} (\bibinfo {year}
  {1999})}\BibitemShut {NoStop}%
\bibitem [{\citenamefont {Neff}\ \emph {et~al.}(2015)\citenamefont {Neff},
  \citenamefont {Feldmeier}, \citenamefont {Horiuchi},\ and\ \citenamefont
  {Weber}}]{src15}%
  \BibitemOpen
  \bibfield  {author} {\bibinfo {author} {\bibfnamefont {T.}~\bibnamefont
  {Neff}}, \bibinfo {author} {\bibfnamefont {H.}~\bibnamefont {Feldmeier}},
  \bibinfo {author} {\bibfnamefont {W.}~\bibnamefont {Horiuchi}}, \ and\
  \bibinfo {author} {\bibfnamefont {D.}~\bibnamefont {Weber}},\ }\href@noop {}
  {\  (\bibinfo {year} {2015})},\ \Eprint {http://arxiv.org/abs/1503.06122}
  {arXiv:1503.06122 [nucl-th]} \BibitemShut {NoStop}%
\bibitem [{\citenamefont {Alvioli}\ \emph
  {et~al.}(2013{\natexlab{b}})\citenamefont {Alvioli}, \citenamefont
  {Ciofi~degli Atti}, \citenamefont {Kaptari}, \citenamefont {Mezzetti},\ and\
  \citenamefont {Morita}}]{alvioli13}%
  \BibitemOpen
  \bibfield  {author} {\bibinfo {author} {\bibfnamefont {M.}~\bibnamefont
  {Alvioli}}, \bibinfo {author} {\bibfnamefont {C.}~\bibnamefont {Ciofi~degli
  Atti}}, \bibinfo {author} {\bibfnamefont {L.~P.}\ \bibnamefont {Kaptari}},
  \bibinfo {author} {\bibfnamefont {C.~B.}\ \bibnamefont {Mezzetti}}, \ and\
  \bibinfo {author} {\bibfnamefont {H.}~\bibnamefont {Morita}},\ }\href
  {\doibase 10.1103/PhysRevC.87.034603} {\bibfield  {journal} {\bibinfo
  {journal} {Phys. Rev. C}\ }\textbf {\bibinfo {volume} {87}},\ \bibinfo
  {pages} {034603} (\bibinfo {year} {2013}{\natexlab{b}})}\BibitemShut
  {NoStop}%
\bibitem [{\citenamefont {Wendt}\ \emph {et~al.}(2014)\citenamefont {Wendt},
  \citenamefont {Carlsson},\ and\ \citenamefont {Ekstr{\"o}m}}]{wendt14}%
  \BibitemOpen
  \bibfield  {author} {\bibinfo {author} {\bibfnamefont {K.}~\bibnamefont
  {Wendt}}, \bibinfo {author} {\bibfnamefont {B.}~\bibnamefont {Carlsson}}, \
  and\ \bibinfo {author} {\bibfnamefont {A.}~\bibnamefont {Ekstr{\"o}m}},\
  }\href@noop {} {\  (\bibinfo {year} {2014})},\ \Eprint
  {http://arxiv.org/abs/1410.0646} {arXiv:1410.0646 [nucl-th]} \BibitemShut
  {NoStop}%
\bibitem [{\citenamefont {Epelbaum}\ \emph {et~al.}(2014)\citenamefont
  {Epelbaum}, \citenamefont {Krebs},\ and\ \citenamefont
  {Mei{\ss}ner}}]{epelbaum14}%
  \BibitemOpen
  \bibfield  {author} {\bibinfo {author} {\bibfnamefont {E.}~\bibnamefont
  {Epelbaum}}, \bibinfo {author} {\bibfnamefont {H.}~\bibnamefont {Krebs}}, \
  and\ \bibinfo {author} {\bibfnamefont {U.~G.}\ \bibnamefont {Mei{\ss}ner}},\
  }\href@noop {} {\  (\bibinfo {year} {2014})},\ \Eprint
  {http://arxiv.org/abs/1412.4623} {arXiv:1412.4623 [nucl-th]} \BibitemShut
  {NoStop}%
\bibitem [{\citenamefont {Sargsian}(2014)}]{sargsian14}%
  \BibitemOpen
  \bibfield  {author} {\bibinfo {author} {\bibfnamefont {M.~M.}\ \bibnamefont
  {Sargsian}},\ }\href {\doibase 10.1103/PhysRevC.89.034305} {\bibfield
  {journal} {\bibinfo  {journal} {Phys. Rev. C}\ }\textbf {\bibinfo {volume}
  {89}},\ \bibinfo {pages} {034305} (\bibinfo {year} {2014})}\BibitemShut
  {NoStop}%
\bibitem [{\citenamefont {Hen}\ \emph {et~al.}(2014)\citenamefont {Hen} \emph
  {et~al.}}]{hen14}%
  \BibitemOpen
  \bibfield  {author} {\bibinfo {author} {\bibfnamefont {O.}~\bibnamefont
  {Hen}} \emph {et~al.} (\bibinfo {collaboration} {Jefferson Lab CLAS
  Collaboration}),\ }\href {\doibase 10.1126/science.1256785} {\bibfield
  {journal} {\bibinfo  {journal} {Science}\ }\textbf {\bibinfo {volume}
  {346}},\ \bibinfo {pages} {614} (\bibinfo {year} {2014})}\BibitemShut
  {NoStop}%
\bibitem [{\citenamefont {Kamuntavi\v{c}ius}\ \emph {et~al.}(2001)\citenamefont
  {Kamuntavi\v{c}ius}, \citenamefont {Kalinauskas}, \citenamefont {Barrett},
  \citenamefont {Mickevi\v{c}ius},\ and\ \citenamefont
  {Germanas}}]{kamuntavicius01}%
  \BibitemOpen
  \bibfield  {author} {\bibinfo {author} {\bibfnamefont {G.}~\bibnamefont
  {Kamuntavi\v{c}ius}}, \bibinfo {author} {\bibfnamefont {R.}~\bibnamefont
  {Kalinauskas}}, \bibinfo {author} {\bibfnamefont {B.}~\bibnamefont
  {Barrett}}, \bibinfo {author} {\bibfnamefont {S.}~\bibnamefont
  {Mickevi\v{c}ius}}, \ and\ \bibinfo {author} {\bibfnamefont {D.}~\bibnamefont
  {Germanas}},\ }\href {\doibase 10.1016/S0375-9474(01)01101-0} {\bibfield
  {journal} {\bibinfo  {journal} {Nucl. Phys. A}\ }\textbf {\bibinfo {volume}
  {695}},\ \bibinfo {pages} {191 } (\bibinfo {year} {2001})}\BibitemShut
  {NoStop}%
\bibitem [{\citenamefont {Navr\'atil}(2004)}]{navratil04}%
  \BibitemOpen
  \bibfield  {author} {\bibinfo {author} {\bibfnamefont {P.}~\bibnamefont
  {Navr\'atil}},\ }\href {\doibase 10.1103/PhysRevC.70.014317} {\bibfield
  {journal} {\bibinfo  {journal} {Phys. Rev. C}\ }\textbf {\bibinfo {volume}
  {70}},\ \bibinfo {pages} {014317} (\bibinfo {year} {2004})}\BibitemShut
  {NoStop}%
\bibitem [{\citenamefont {Navr\'atil}\ \emph {et~al.}(2000)\citenamefont
  {Navr\'atil}, \citenamefont {Kamuntavi\v{c}ius},\ and\ \citenamefont
  {Barrett}}]{navratil00}%
  \BibitemOpen
  \bibfield  {author} {\bibinfo {author} {\bibfnamefont {P.}~\bibnamefont
  {Navr\'atil}}, \bibinfo {author} {\bibfnamefont {G.~P.}\ \bibnamefont
  {Kamuntavi\v{c}ius}}, \ and\ \bibinfo {author} {\bibfnamefont {B.~R.}\
  \bibnamefont {Barrett}},\ }\href {\doibase 10.1103/PhysRevC.61.044001}
  {\bibfield  {journal} {\bibinfo  {journal} {Phys. Rev. C}\ }\textbf {\bibinfo
  {volume} {61}},\ \bibinfo {pages} {044001} (\bibinfo {year}
  {2000})}\BibitemShut {NoStop}%
\end{thebibliography}%

\end{document}